%% file: main.tex
\documentclass[runningheads,orivec]{llncs}

\usepackage[T1]{fontenc}
\usepackage{graphicx}
\usepackage{amsmath}
\usepackage{amsfonts}
\usepackage{amssymb}
\usepackage{xfrac}
\usepackage{multirow}
\usepackage[rgb,dvipsnames,svgnames,x11names]{xcolor}
\usepackage{xurl}
\usepackage{tabularx}
\usepackage{enumerate}
\usepackage{colortbl}
\usepackage[USenglish]{babel}
\usepackage{orcidlink}
\usepackage{tikz}
\usetikzlibrary{positioning}
\usetikzlibrary{calc}
\usetikzlibrary{ext.paths.ortho}
\usetikzlibrary{shapes.geometric}
\usetikzlibrary{intersections}
\usetikzlibrary{patterns}

\usepackage{color}

\usepackage{hyperref}
\usepackage[acronym,hyperfirst=false]{glossaries-extra}
\newignoredglossary{ignored}
\defglsentryfmt[ignored]{\GlsUseAcrEntryDispStyle{dua-desc}}
\input{acronyms.tex}
\makenoidxglossaries

\usepackage[nameinlink]{cleveref}
\crefname{section}{Sec.}{Sec.}
\crefname{subsection}{Sec.}{Sec.}
\crefname{subsubsection}{Sec.}{Sec.}
\crefname{figure}{Fig.}{Figs.}
\crefname{table}{Tab.}{Tabs.}

\usepackage{pdflscape}

\makeatletter
\newcommand{\printfnsymbol}[1]{%
  \textsuperscript{\@fnsymbol{#1}}%
}
\makeatother

\input{figures}

\begin{document}
\title{Systematic Cybersecurity Risk Analysis of European Rail Traffic Management System}
\titlerunning{Systematic Cybersecurity Risk Analysis of ERTMS}

\author{
Kacper Darowski\thanks{These authors contributed equally to this work. \\
This is the \textit{extended version} of the paper accepted at ARES 2026 CPRA. }\inst{1}\orcidlink{0000-0001-5924-6484}
\and 
Sebastian N. Peters\printfnsymbol{1}\inst{1,2}\orcidlink{0009-0007-6421-4023}
\and
Lukas Lautenschlager\printfnsymbol{1}\inst{1,2}\orcidlink{0009-0002-0418-9819}
}
\authorrunning{Darowski, Peters, and Lautenschlager}

\institute{
TUM CIT, Technical University of Munich, Germany\\
\and
Fraunhofer AISEC, Garching bei München, Germany\\
\email{\{sebastian.peters,lukas.lautenschlager\}@aisec.fraunhofer.de}
}
\maketitle 

\begin{abstract}
\gls{ertms} is a widely adopted standard unifying train management in the EU. 
While the standard allows for use cases like fully autonomous driving, cybersecurity has been an afterthought. 
Risk analysis enables the systematic assessment and prioritization of threats and mitigations. 
To date, it remains unclear which threats are most significant in \gls{ertms}. 
This study systematically models components of \gls{ertms} and analyzes their security in light of threats identified in the underlying technologies. 
The results suggest a concerning state of \gls{ertms}, despite its critical role in railway safety.
The use of legacy standards like EuroBalises and \gls{gsmr} introduces vulnerabilities that persist across minimal \gls{ertms} implementations, deployments incorporating various optional safety measures, and prospective future evolutions of the system, e.g., adopting \gls{frmcs}. 
Fully transitioning to \gls{etcs} level 2 was identified as the most significant improvement to \gls{ertms} cybersecurity. 
The results indicate that a shift of \gls{ertms} toward security is required to ensure availability and safe operation. 
While the chosen methodology proved its feasibility and shows remaining weaknesses of \gls{ertms}, future work is needed to develop railway-centric adaptations to improve the quantification and evaluation of the computed risks. 


\end{abstract}

\glsresetall
\section{Introduction}
To promote the development and operation of international railways, the EU introduced the \gls{ertms} \cite{96-40-ec} to establish interoperable signaling, communication, and train management systems.

\gls{ertms} enables a number of technologies like  standardized signaling, \gls{ota} communication, and \gls{ato}.
Having first become mandatory for new high-speed networks in 2002 \cite{2002-731-ec}, and for regular railways in 2004 \cite{2004-49-ec}, \gls{ertms} has seen mixed adoption rates ranging from full deployment in Belgium and Luxembourg to a coverage of less than 1\% in Germany \cite{eu-rail-infrastructure-state-2025}. 

\gls{ertms} has even expanded beyond the borders of the EU to countries like Australia or Thailand\footnote{see \url{https://www.ertms.net/deployment-world-map/}, Accessed: 2026-06-08}, indicating global trust in the responsible \gls{era}.
The \gls{era} published the first specification draft in 1996 \cite{96-40-ec} and subsequently updated it, most recently as ``Baseline 4'' in 2023 \cite{l-2023-222}. 
Nevertheless, this long legacy raises the question to what extent cybersecurity has been part of the standardization process.

In the past decade, various attacks on railway infrastructure have occurred across multiple countries, typically in the form of \gls{dos} or the release of personal information \cite{cybersecurity-railway-review-2022}. 
Legacy cyber-physical systems, in particular, are vulnerable to low-cost attacks, as was the case with Polish trains being remotely halted by broadcasting a stop command over the radio in 2023 \cite{train-brakes-radio-2025}. 
Apart from attacks carried out for personal gain or targeted service disruptions, concerns about exploitation in military warfare are on the rise amidst heightened geopolitical tensions. 
While the EU migrating its signaling and train management to highly computerized systems is a necessary step in expanding and unifying its rail networks, the relatively dated nature of \gls{ertms}, compared to the rapid pace of cybersecurity developments, calls for a closer investigation of the status quo.
The contributions of this paper are:
\begin{enumerate}[\bf{C}1]
    \item We adapt and apply a modular cybersecurity risk assessment approach to systematically model and evaluate \gls{ertms}.
    \item We derive/compare the cybersecurity risk profiles of current/future \gls{ertms} configurations, identifying the assets that expose the largest attack surface.
\end{enumerate}

We summarize related research in \Cref{sec:related} and explain the core concepts of \gls{ertms} in \Cref{sec:background}, which are relevant to the system model (\cref{sec:model}) and risk analysis (\cref{sec:evaluation}). 
\Cref{sec:methodology} describes our approach to modeling and analyzing \gls{ertms}. 
The identified threats and controls are described in \Cref{sec:controls}. 
The results of our analysis are shown in \Cref{sec:evaluation} and discussed in \Cref{sec:discussion}. 
\Cref{sec:conclusion} concludes our findings. 

\section{Related Work}
\label{sec:related}
This section summarizes the existing literature using the following trichotomy:
cybersecurity of generic railway infrastructure, of \gls{ertms}, and risk analysis.

\textbf{Railway Security:}
Academic and industrial work agrees that the growing digitalization and connectivity of railway \gls{ot} increases the cyberattack surface of a strategically critical infrastructure \cite{signalling-cybersecurity-mission-centric-2016,railway-cybersecurity-strategy-2024,railway-cybersecurity-review-2022,railway-cybersecurity-overview-2024}. 
Surveys highlight safety-security co-engineering, simulation and modeling, and the use of \gls{ai} and blockchain as main research trends, while pointing to gaps in validation, \gls{lte}/5G security, cybersecurity implementation, and quantitative risk management \cite{smart-railways-cybersecurity-data-communication-2025,railway-cybersecurity-review-2022}. 
Technical contributions focus on network-level protections such as segregation via firewalls/gateways, \glspl{ids}, \gls{dpi}, \gls{ai}-based detection, and strong encryption, but also stress practical constraints arising from long asset lifetimes, geographical dispersion, and vendor lock-in, and therefore advocate risk-based and cost-aware security engineering \cite{smart-railways-cybersecurity-data-communication-2025,railway-cybersecurity-strategy-2024,cyberbezpieczenstwo-sterowanie-2020,cyberbezpieczenstwo-kierowanie-sterowanie-2017,railway-cybersecurity-overview-2024}. 
At the normative level, CENELEC TS~50701 and the recent EU Rail System-Pillar specifications provide life-cycle guidance, generic threat scenarios, and baseline cybersecurity requirements for signaling and interlocking systems \cite{interlocking-interoperability-cybersecurity-requirements-2020,era_com,era_comp,era_spr,era_scss,railway-cybersecurity-strategy-2024,railway-cybersecurity-overview-2024}. 
Our work complements this horizontal view with an independent, technology-aware risk analysis, including explicit attack trees and detailed assessments of \gls{frmcs} and \gls{ato}. 

\textbf{ERTMS Security:}
Cybersecurity analyses of \gls{ertms} concentrate mainly on its communication and cryptographic mechanisms. Early work on EuroRadio's ``Safety Layer'' over \gls{gsmr} exposed weak cryptographic primitives (3DES) and problematic key-management practices, and suggested more robust designs \cite{related-ertms-2015-lopez}. 
Subsequent studies analyzed denial-of-service and jamming against \gls{gsmr}, resource-exhaustion and \gls{prng} attacks, and again questioned key management, while calling for systematic vulnerability assessments \cite{related-ertms-2018-gabriel,related-ertms-2019-kochan}. 
A second line of work focuses on EuroBalises: simulation-based studies showed that missing integrity protection and remotely modifiable firmware can enable manipulation of automatic train halting; countermeasures include retrofitted authentication tags and anomaly detection \cite{related-ertms-2019-lim}. 
Tests in related \gls{cbtc} systems confirmed that low-cost equipment suffices to jam balise telegrams and also highlighted risks in Wi-Fi-based intra-vehicular communication \cite{related-ertms-2023-soderi}. 
More recent contributions discuss system-level mitigations for future \gls{ertms} evolutions, in particular \gls{frmcs}, recommending centralized ``overarching security mechanisms'' (such as \glspl{soc}, \gls{mdm}, and a sector-specific \gls{pki} and \gls{dns}) as well as cryptographic agility, continuous mediation, and \gls{ai}-based observability and “sustained visibility” for already deployed assets \cite{related-ertms-2025-benjumea,related-ertms-2022-koop}. 
These works mostly address individual components or mechanisms; in contrast, we analyze the complete \gls{ertms}, from legacy \gls{gsmr} and EuroBalises to \gls{frmcs} and \gls{ato}, within a unified risk framework. 

\textbf{Cybersecurity Risk Analysis:}
Methodologically, our work builds on established security risk-assessment approaches. 
In the automotive domain, Wolf et al.\ derive risk levels by combining \gls{cem}-style attack potentials with damage estimates based on adjusted \gls{sil} levels and attack trees \cite{cem-2022,related-risk-2012-wolf}. 
Eichler et al.\ generalize this into the \gls{mora} framework, which supports scalable, function-oriented, and technology-aware analyses \cite{related-risk-2015-eichler-mora}.
It has been applied before to similar complex use cases, such as autonomous driving \cite{wagner_Cybersecurity_2023}. 
For railways, prior work includes safety-focused risk models \cite{related-risk-2017-leitner} and approaches that jointly consider safety and security interactions \cite{related-risk-2021-aktouche}. 
Only a few studies apply cybersecurity risk assessment specifically to \gls{ertms}: 
Bloomfield et al.\ sketch a methodology for \gls{ertms}-based systems but disclose limited results, while other work concentrates on single mechanisms such as 3DES in EuroRadio or secure \gls{ertms} architectures and key management \cite{bloomfield_Risk_2016,pepin_Risk_2016,Thomas2019ASD}. 
Existing literature thus either remains high-level or focuses on narrowly scoped technical issues. 
We close this gap by applying \gls{mora} to the full \gls{ertms} specification, including new standardizations such as \gls{frmcs}, and by publishing explicit attack trees and detailed, system-wide risk profiles instead of only presenting a methodology.

\section{Background}
\label{sec:background}
This section explains the current state of \gls{ertms} standards in the scope of this analysis. It introduces the components, interfaces, and cryptographic mechanisms that later form the system model in \cref{sec:model}.

\textbf{European Train Control System:} \label{sec:background-etcs}
Within \gls{ertms}, the \gls{etcs} signaling and control system has the most direct impact on passenger safety. It ensures that no two trains enter the same track segment simultaneously; we consider specification version~4.0.0 \cite{subset026}. 
\gls{etcs} defines four operational levels: level~0 (no \gls{etcs} on the \gls{ts}), level ``\gls{ntc}'' (legacy systems connected via a \gls{stm}), level~1 (\gls{etcs} train supervision with data from EuroBalises, EuroLoops, \glspl{rbc}, and \glspl{riu}), and level~2 (continuous radio communication with \glspl{rbc}, with EuroBalises mainly providing positional data; EuroLoops and \glspl{riu} are not used) \cite{subset036,subset044,subset026}. 
EuroBalises and EuroLoops are spot and semi‑continuous track devices that transmit wayside information to the \gls{obu} via inductive coupling, while \glspl{rbc} and \glspl{riu} provide radio-based ``infill'' information \cite{subset036,subset044,subset026}. 
The pivotal datum is the \gls{ma}, which authorizes a train to proceed up to an \gls{eoa} or \gls{loa}; \glspl{ma} can be extended but not revoked, only cooperatively shortened \cite{subset026}. Additional data includes emergency stops, temporary speed restrictions, gradients, and further track conditions. 
Balise linking triggers an emergency halt if expected balises or loops are not detected.
This may be suppressed near known interference sources via \gls{bmm} and \gls{vbc} \cite{subset125,subset026}. 
The \gls{etcs} \gls{obu} supervises speed and can apply brakes but not traction, which remains with the driver or \gls{ato}. 
Position is computed mainly from odometry referenced to balises, with \glspl{gnss} used only as time sources \cite{subset147,subset026}. 

\textbf{GSM-R and FRMCS:}\label{sec:background-gsmr}
Today, radio communication between train and \glspl{rbc}/\glspl{riu} uses \gls{gsmr}, a dedicated mobile network operating in circuit- and packet-switched modes; packet mode relies on standard \gls{tcp}/\gls{ip}, with the EuroRadio module resolving \gls{rbc}/\gls{riu} addresses via \gls{dns} \cite{subset037}. 
A ``Safety Layer'' supplements GSM with traffic encryption and integrity protection using three symmetric keys: \texttt{KTRANS} (transport), \texttt{KMAC} (authentication) and \texttt{KSMAC} (session).
Fresh \texttt{KSMAC} keys are derived from \texttt{KMAC} and nonces via \gls{3des}, and a 64-bit \gls{cbc}-\gls{mac} provides integrity \cite{subset037}. 
The upcoming \gls{frmcs} replaces \gls{gsmr} with a 5G-based system that offers inherent authentication and encryption, plus TLS-protected end-to-end communication. 
\gls{dns} is again used for \gls{ip} resolution \cite{subset037,at7800}.

\textbf{Automatic Train Operation:} \label{sec:background-ato}
\gls{ato} is an optional subsystem enabling (semi-)autonomous driving and automatic station stops. 
The \gls{ato} \gls{obu} receives its data exclusively over \gls{frmcs} from the trackside \gls{ato} system.
\glspl{sp} describe infrastructure segments (stop points, speed profiles, tunnels, platforms), while a \gls{jp} combines multiple \glspl{sp} with additional infrastructure and timetable constraints. 
Internally, the \gls{ato} \gls{obu} contains three modules: \gls{ttsm}, \gls{ssem} (respecting \gls{etcs} speed supervision), and \gls{atsm} (accurate stopping at defined points). 
Their speed recommendations are combined by taking the minimum, and \gls{ato} directly controls both traction and brakes, while also having the option to manage door opening/closing according to dwell times and platform data \cite{subset125}. 
\gls{etcs} remains the safety authority and can override \gls{ato} via braking.

\textbf{Key Management:} \label{sec:background-kmc}
ERTMS defines a distributed key management architecture for \texttt{KTRANS} and \texttt{KMAC} \cite{subset038,subset137}. 
Each entity belongs to a key management domain served by a \gls{kmc}, which generates and distributes \texttt{KMAC} and handles inter-domain requests. 
Keys can be provisioned via an off-line, out-of-band interface (using \gls{3des} and a \gls{cbc}-\gls{mac}, with unclear initial secret installation) \cite{subset038} or via an on-line TLS-protected channel that relies on \texttt{KTRANS} and a \gls{pki} for certificate-based distribution \cite{subset146,subset137,subset026}. 
With full transition to \gls{frmcs}, legacy \texttt{KMAC}/\texttt{KSMAC} usage is phased out and all cryptographic material is to be managed by a sector-wide \gls{pki} \cite{subset146}.

\section{Methodology}
\label{sec:methodology}
The methodology follows four basic steps: 

\begin{enumerate}
   \item \textbf{Literature Review:} Identify relevant specification documents and survey existing research on \gls{ertms}, to obtain a detailed understanding.
    \item \textbf{System Modeling:} Construct an abstract model of the \gls{ertms}.
   \item \textbf{Risk Analysis:} Perform a risk analysis of the derived system.
   \item \textbf{Validation:} Compare the results with related work.
\end{enumerate}

The majority of the \gls{ertms} specification is publicly accessible on \gls{era}'s website. 
It is split into multiple documents called ``subsets''. 
Due to the large number of these subsets, compiling the information they contain requires substantial effort in cross-referencing the documents.
After careful consideration, the previously described \gls{mora} framework has been chosen for the risk analysis for the following reasons: 
Its \textbf{iterative and modular nature} is well-suited for the scale of \gls{ertms}, as granularity can be adapted based on early findings. 
Establishing the \gls{toe} ensures \textbf{completeness} with respect to the selected model and threats, minimizing subjectivity and bias (though not guaranteeing completeness for the real-world system, which depends on the analyst's abstractions). 
To tackle the scope of the specification in the risk analysis, the following approach to \gls{mora} was chosen:
\begin{enumerate}
    \item \textbf{Creation of \gls{toe}}: The system model is formalized into the \gls{toe} by identifying system components, connections, and used technologies. Data flows for each connection are modeled at the chosen level of granularity. 
    \item \textbf{Function Assignment}: The data is grouped into ``functions'' based on their purposes and effects. Any datum can be part of multiple functions. Based on the established data flows, the components and connections are automatically mapped to the same functions.
    \item \textbf{Quantification of Security Goals}: For each function, the impairment of \textit{Confidentiality}, \textit{Integrity}, and \textit{Availability} is assessed across four damage classes aligned with ISO/SAE 21434 \cite{ISO_21434}: \textit{Safety}, \textit{Financial}, \textit{Legal (also covering Privacy)}, and \textit{Operational}. Each combination of function and impairment constitutes a security goal. Following \gls{mora}, damages are assessed per function at worst-case severity, as any compromise of a given security goal produces the same downstream consequences regardless of the specific attack vector.
    \item \textbf{Threat Establishment}: Threats are identified using a subset of \gls{stride} (spoofing, tampering, information disclosure, and \gls{dos}; repudiation and elevation of privileges are subsumed by spoofing and tampering as \gls{ertms} relies predominantly on physical domain separation, and forged messages inherently deny authentic origin) applied to components or connections. Each threat is rated with a \gls{rap} following \gls{cem} \cite{cem-2022} across five categories: \textit{Elapsed Time}, \textit{Required Expertise}, \textit{Knowledge of \gls{toe}}, \textit{Window of Opportunity}, and \textit{Required Equipment}. According to \gls{cem} \cite{cem-2022}, the \gls{rap} levels stem from summed category scores: 0\textendash 9 (Basic), 10\textendash 13 (Enhanced-Basic), 14\textendash 19 (Moderate), 20\textendash 24 (High), and $\geq$25 (Beyond High).
    \item \textbf{Defining Controls}: Different controls that can secure the system are identified through an iterative process. We begin with the set of controls stemming directly from the \gls{ertms} specification and our literature analysis. For each unaddressed threat, further literature is reviewed to identify applicable countermeasures. Next, we identify threats that render some control ineffective; e.g., gaining access to a private key renders a secure channel protected by that key ineffective. Controls beyond those steps are not considered.
    \item \textbf{Compiling Attack Trees}: For each threat, an attack tree is built recursively. The root represents the threat, and children are ``preparation threats'' weakening controls protecting against it. Subtrees that share the same root are duplicated, while paths whose \gls{rap} is fully dominated by a shorter path are pruned.
    \item \textbf{Risk Estimation}: The combined required attack potential of each of the attack trees is calculated by summing each path and multiplying it by the damage levels of the broken security goals, yielding a ``risk level'' for the tree.
    \item \textbf{Evaluation of Controls}: Finally, the risks are recalculated for different sets of active controls, to gain an understanding of the effectiveness of various countermeasures. This can then be compared with the claims made by the designers of these controls and other researchers.
\end{enumerate}

\section{System Model}
\label{sec:model}
This section describes the developed system model as formalized in \gls{mora}'s \gls{toe}. The identified components can be segregated into four broader categories.

At the core of the \gls{ob} we find the \gls{etcs} \gls{obu}, which interfaces with all the other components within a train. For our purposes, we consider the \gls{dmi}, the \gls{ato} \gls{obu}, a representative \gls{stm}, the \gls{ord}, the EuroRadio module, and the vehicle interface \cite{subset026}. Some of the abstraction choices include omitting distinct communication modules for EuroBalises and EuroLoops and merging the \gls{frmcs} functionality into EuroRadio.
All \gls{ob} components are connected to the \gls{etcs} \gls{obu} via a physical Ethernet network running PROFINET \cite{subset147,subset026}. Historically, \gls{mvb} and \gls{can} have been used, but we ignore this in our model, as they are no longer permitted in new vehicles. Additionally, the \gls{ato} \gls{obu} has its own Ethernet connections to the EuroRadio, the \gls{dmi}, and the vehicle interface (possibly multiple, but this distinction is irrelevant to our analysis).

Our model's \gls{ts} comprises four train communication systems: 
the EuroBalise, the EuroLoop, the \gls{rbc}, and the \gls{riu}. It is sufficient to model only one instance of each component. We choose not to model any further components, like \glspl{leu}, \gls{ntc} \gls{ts}, or the interlocking backbone, as these are beyond the scope of \gls{ertms}. For the same reason, we do not model any internal connections on the \gls{ts} \cite{subset026}.
The \gls{ts} interfaces transmit messages to the \gls{ob} via four channels: 
induction between the EuroBalise and the \gls{etcs} \gls{obu}, respectively, the EuroLoop, \gls{gsmr} between the \gls{rbc} and the EuroRadio, as well as an equivalent connection with the \gls{riu}. 
We ignore \gls{frmcs} here, and view it as a control instead (see \Cref{sec:controls}) \cite{subset026}.

Some services in \gls{ertms} are accessed via the Internet: 
\gls{pki}, \gls{dns} resolver, and \gls{ato} \gls{ts}. 
We also include a gateway connecting the \gls{gsmr} network to the Internet. 
We form one \gls{tcp} connection from the \gls{pki} to the gateway, the \gls{rbc}, the \gls{riu}, and the \gls{ato} \gls{ts}. The \gls{dns} resolver connects to the gateway via \gls{udp}. The train's Internet connectivity is modeled as a \gls{gsmr} connection between the EuroRadio and the gateway \cite{subset125,subset146,subset037}. 

We define two components relevant to \gls{km}: a local and an external \gls{kmc}. The local \gls{kmc} is connected redundantly via \gls{tcp} and \gls{oob} to the \gls{rbc}, the \gls{riu}, and the external \gls{kmc}. 
Both \glspl{kmc} interface with the \gls{pki} over \gls{tcp} \cite{subset114,subset038,subset137}.

\begin{figure}[!tb]
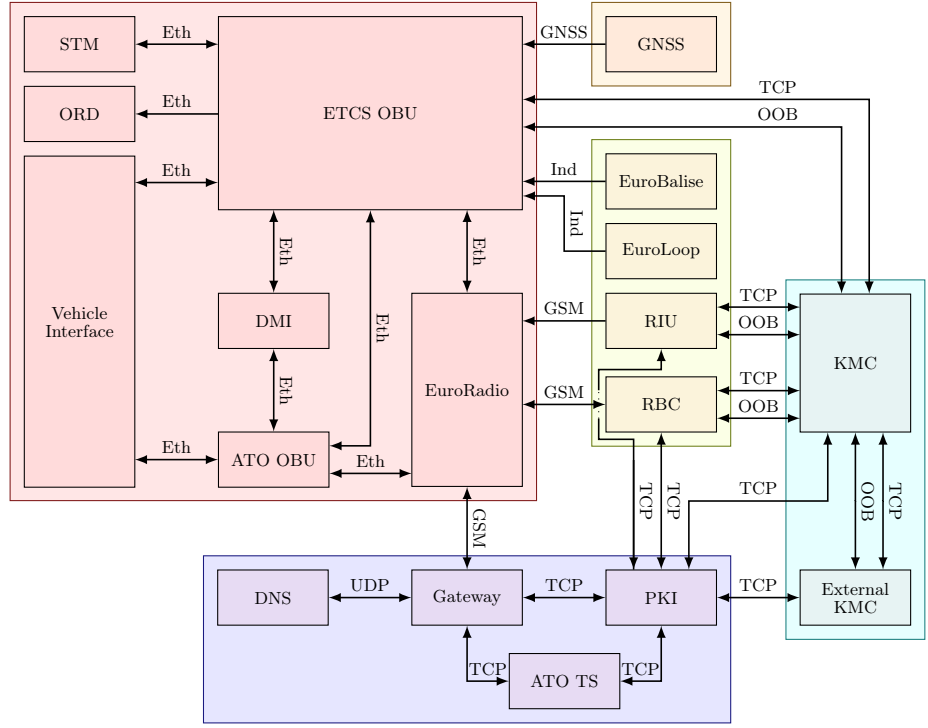

    \centering
    \figureModel
  \caption{System components grouped by domain (\gls{ob}, \gls{ts}, Internet, \gls{km}) and their connections. Edge labels denote the communication technology: Eth = Ethernet, Ind = Magnetic induction, \gls{gsm} = \gls{gsmr}, \gls{oob} = Out-of-Band}
  \label{fig:model}
\end{figure}

Lastly, a \gls{gnss} component with an interface to the \gls{etcs} \gls{obu} is included for some configurations. 
\Cref{fig:model} illustrates all components and their connections.

The data units transferred on each connection were extracted from the relevant specification documents \cite{subset139,subset146,subset037,subset027,subset114,subset038,subset137,subset035,subset026,subset034,eraertms015560}. We omit listing them and instead focus on the identified functions; we define one function related to traction control, one to brake systems, one to \glspl{ma}, two to the \gls{dmi}, three to various status reports, five to different track information, one to key management and one to certificate management, three to mode and level transitions, one for door control, as well as one for odometry, current time, version negotiation, communication establishment and termination, logging, acknowledgements, \gls{eoi}, error reporting, safe radio supervision, \gls{vbc} and balise linking, emergency stops, and \gls{ntc} each, for a total of $37$ functions.

\newcolumntype{C}[1]{>{\centering\arraybackslash}p{#1}}

\definecolor{veryhigh}{Hsb}{0,1,0.80}
\definecolor{high}{Hsb}{37,0.85,.85}
\definecolor{moderate}{Hsb}{52,.85,.87}
\definecolor{low}{Hsb}{85,.70,.93}
\definecolor{vlow}{Hsb}{180,.3,1}

\definecolor{none}{RGB}{200, 200, 200}
\definecolor{brkcon}{Hsb}{270,.5,.85}

\newcommand{\vhigh}{\cellcolor{veryhigh} VH}
\newcommand{\high}{\cellcolor{high} H}
\newcommand{\med}{\cellcolor{moderate} M}
\newcommand{\low}{\cellcolor{low} L}
\newcommand{\vlow}{\cellcolor{vlow} VL}
\newlength{\ciacellw}
\setlength{\ciacellw}{0.5cm}

\newcommand{\nmark}{%
  \strut\smash{\makebox[0pt][l]{%
    \hspace*{\dimexpr-0.5\ciacellw-\tabcolsep\relax}%
    \raisebox{-\dp\strutbox}[0pt][0pt]{%
      \tikz \fill[pattern=north east lines, pattern color=black, line width=2pt] 
        (0,0) rectangle (\dimexpr\ciacellw+2\tabcolsep\relax, \ht\strutbox+\dp\strutbox);%
    }%
  }}%
  \makebox[0pt][c]{-}
}

\newcommand{\legendsquare}[2][0.08em]{
  \def\sqsize{0.7em}%
  \rlap{\rule{\sqsize}{\sqsize}}%
  \hspace{#1}%
  \raisebox{#1}{\textcolor{#2}{\rule{\dimexpr\sqsize-2\dimexpr#1\relax\relax}{\dimexpr\sqsize-2\dimexpr#1\relax\relax}}}%
  \hspace{#1}%
}

\newcommand{\legendsquarepattern}[2][0.08em]{
  \def\sqsize{0.7em}%
  \rlap{\rule{\sqsize}{\sqsize}}%
  \hspace{#1}%
  \raisebox{#1}{%
    \tikz \fill[white] (0,0) rectangle 
      (\dimexpr\sqsize-2\dimexpr#1\relax\relax, \dimexpr\sqsize-2\dimexpr#1\relax\relax);%
    \hspace{-\dimexpr\sqsize-2\dimexpr#1\relax\relax}%
    \tikz \fill[pattern=#2, pattern color=black] (0,0) rectangle 
      (\dimexpr\sqsize-2\dimexpr#1\relax\relax, \dimexpr\sqsize-2\dimexpr#1\relax\relax);%
  }%
  \hspace{#1}%
}

The damage to each function is scored in four categories, each with five levels. The four categories are Safety, Financial, Legal (including Privacy-related damages), and Operational damages. Each category (with some exceptions) contains multiple levels with thresholds ranging from \textit{Very Low} to \textit{Very High}. Table \ref{tab:damage_thresh} shows the thresholds of each damage category paired with each damage level.

\begin{table}[!tb]
    \centering
    \caption{Overview of different damage classes (\textit{Safety}, \textit{Financial}, \textit{Legal (including privacy laws)}, \textit{Operational}) with their severities (\legendsquare{veryhigh}\textit{\underline{V}ery \underline{H}igh}, \legendsquare{high}\textit{\underline{H}igh}, \legendsquare{moderate}\textit{\underline{M}oderate}, \\ \legendsquare{low}\textit{\underline{L}ow}, \legendsquare{vlow}\textit{\underline{V}ery \underline{L}ow}) and their corresponding stakeholder and criteria. For example, the safety category is divided into different severities based on threshold values of \glsxtrlong{fwi} (\gls{fwi}), a standard metric to measure safety in railway systems \cite{lin_International_2023}.}
    \label{tab:damage_thresh}
    \begin{tabular}{|c|c|c|p{5cm}|}
        \hline
        \textbf{Damage class} & \textbf{Damage level} & \textbf{Stakeholder} & \textbf{Description} \\
        \hline\hline
        Safety & \low & Customer & $\gls{fwi} < 0.1$ \\
        \hline
        Safety & \med & Customer & $0.1 \leq \gls{fwi} < 1$\\
        \hline
        Safety & \high & Customer & $1 \leq \gls{fwi} < 3$\\
        \hline
        Safety & \vhigh & Customer & $3 \leq \gls{fwi}$ or at least $1$ death\\
        \hline\hline
        Financial & \vlow & Manufacturer & No or tolerable financial damage\\
        \hline
        Financial & \low & Manufacturer & Undesirable damage: Single unit of rolling stock or infrastructure very limited damaged \\
        \hline
        Financial & \med & Manufacturer & Moderate damage: Limited amount of rolling stock or limited amount of infrastructure damaged \\
        \hline
        Financial & \high & Manufacturer & Substantial damage: Large amount of rolling stock or large damages to infrastructure or buildings \\
        \hline
        Financial & \vhigh & Manufacturer & Existence-threatening damage: Permanent damage to an entire fleet \\
        \hline\hline
        Legal & \high & Manufacturer & Violation of law\\
        \hline
        Legal & \vhigh & Manufacturer & Massive violation of law\\
        \hline\hline
        Operational & \vlow & Customer & Affected passenger comfort or delay below a minute\\
        \hline
        Operational & \low & Customer & Service slightly impaired: Delay below ten minutes\\
        \hline
        Operational & \med & Customer & Service moderately impaired: Large delay or cancellation of a single train \\
        \hline
        Operational & \high & Customer & Service highly impaired: Large delay or cancellation of multiple trains \\
        \hline
        Operational & \vhigh & Customer & Service completely ceased: Entire railway network or large parts of it disabled \\
        \hline
    \end{tabular}
    
\end{table}

The compromise of the majority of these functions' security goals is associated with a \textit{Very High} or \textit{High} damage level, as it may lead to consequences ranging from section blockages to level-crossing accidents resulting in injuries or fatalities. Security goals with \textit{Moderate} and \textit{Low} damage levels relate to moderate train delays, e.g., due to a broken \gls{ma} shortening, or passenger comfort, e.g., due to a compromise of air conditioning. Table \ref{tab:damage_assessment} lists all functions with their security goals and assigned damage severity in each class. For example, the integrity of the function \gls{ma} yields \textit{Very High} safety damages, as authorizing a train to move into an occupied block can lead to collisions between trains. This also applies to the \textit{High} financial damages caused through potential collisions between trains. Operational damages arise from attacks on the integrity or availability of the \gls{ma} function—e.g., spoofing a blocked signal or dropping authorization messages—which can prevent train movement authorizations to free blocks, causing significant delays and network blockages.

\begin{table}[!tb]
    \centering
    \caption{Damage level assessment per function and security goal. Each color represents a severity (\legendsquare{veryhigh}\textit{\underline{V}ery \underline{H}igh}, \legendsquare{high}\textit{\underline{H}igh}, \legendsquare{moderate}\textit{\underline{M}oderate}, \legendsquare{low}\textit{\underline{L}ow}, \legendsquare{vlow}\textit{\underline{V}ery \underline{L}ow}, \legendsquarepattern{north east lines}\textit{No Damages}). An asterisk (*) at the function name indicates that one or more of that function's security goals break a control, e.g., the integrity of the certificate management breaks TLS.}
    \label{tab:damage_assessment}
    \begin{tabular}{|l|C{0.5cm}|C{0.5cm}|C{0.5cm}|C{0.5cm}|C{0.5cm}|C{0.5cm}|C{0.5cm}|C{0.5cm}|C{0.5cm}|C{0.5cm}|C{0.5cm}|C{0.5cm}|}
    \hline
    \multirow{2}{*}{\textbf{Function}} 
        & \multicolumn{3}{c|}{\scriptsize\textbf{Safety}} 
        & \multicolumn{3}{c|}{\scriptsize\textbf{Financial}} 
        & \multicolumn{3}{c|}{\scriptsize\textbf{Legal}} 
        & \multicolumn{3}{c|}{\scriptsize\textbf{Operational}} \\
    \cline{2-13}
    & \scriptsize\textbf{C} & \scriptsize\textbf{I} & \scriptsize\textbf{A} 
    & \scriptsize\textbf{C} & \scriptsize\textbf{I} & \scriptsize\textbf{A} 
    & \scriptsize\textbf{C} & \scriptsize\textbf{I} & \scriptsize\textbf{A} 
    & \scriptsize\textbf{C} & \scriptsize\textbf{I} & \scriptsize\textbf{A} \\
    \hline
    Traction Control 
        & \nmark & \vhigh & \vhigh 
        & \nmark & \high & \high 
        & \nmark & \nmark & \nmark 
        & \nmark & \high & \high \\
    \hline
    Brake Control*
        & \nmark & \vhigh & \vhigh 
        & \nmark & \high & \high 
        & \nmark & \nmark & \nmark 
        & \nmark & \high & \high \\
    \hline
    Door Control 
        & \nmark & \vhigh & \nmark 
        & \nmark & \nmark & \nmark 
        & \nmark & \nmark & \nmark 
        & \nmark & \med & \med \\
    \hline
    Certificate Manag.*
        & \nmark & \nmark & \nmark 
        & \nmark & \nmark & \nmark 
        & \nmark & \nmark & \nmark 
        & \nmark & \high & \high \\
    \hline
    Track Condition 
        & \nmark & \vhigh & \nmark 
        & \nmark & \high & \nmark 
        & \nmark & \nmark & \nmark 
        & \nmark & \high & \high \\
    \hline
    Movement Authority 
        & \nmark & \vhigh & \nmark 
        & \nmark & \high & \nmark 
        & \nmark & \nmark & \nmark 
        & \nmark & \high & \high \\
    \hline
    Balise Linking*
       & \nmark & \nmark & \nmark 
        & \nmark & \nmark & \nmark 
        & \nmark & \nmark & \nmark 
        & \nmark & \nmark & \nmark \\
    \hline
    Version Negotiation 
         & \nmark & \nmark & \nmark 
        & \nmark & \nmark & \nmark 
        & \nmark & \nmark & \nmark 
        & \nmark & \high & \high \\
    \hline
    \gls{ntc} 
       & \nmark & \vhigh & \nmark 
        & \nmark & \high & \nmark 
        & \nmark & \nmark & \nmark 
        & \nmark & \high & \high \\
    \hline
    Display 
        & \nmark & \nmark & \nmark 
        & \nmark & \nmark & \nmark 
        & \nmark & \nmark & \nmark 
        & \nmark & \nmark & \nmark \\
    \hline
    Air Intake 
        & \nmark & \low & \low 
        & \nmark & \nmark & \nmark 
        & \nmark & \nmark & \nmark 
        & \nmark & \vlow & \vlow \\
    \hline
    Power 
        & \nmark & \high & \nmark 
        & \nmark & \med & \low 
        & \nmark & \nmark & \nmark 
        & \nmark & \high & \high \\
    \hline
    Logging 
       & \nmark & \nmark & \nmark 
        & \nmark & \nmark & \nmark 
        & \nmark & \high & \high 
        & \nmark & \nmark & \nmark \\
    \hline
    Level Crossing 
        & \nmark & \vhigh & \vhigh 
        & \nmark & \med & \med 
        & \nmark & \nmark & \nmark
        & \nmark & \high & \high \\
    \hline
    Communication Est. 
        & \nmark & \nmark & \nmark 
        & \nmark & \nmark & \nmark 
        & \nmark & \nmark & \nmark 
        & \nmark & \high & \high \\
    \hline
    Key Management*
         & \nmark & \nmark & \nmark 
        & \nmark & \nmark & \nmark 
        & \nmark & \nmark & \nmark 
        & \nmark & \high & \high \\
    \hline
    Mode Transition 
       & \nmark & \vhigh & \nmark 
        & \nmark & \med & \nmark 
        & \nmark & \nmark & \nmark 
        & \nmark & \high & \high \\
    \hline
    \gls{etcs} Transition 
        & \nmark & \nmark & \nmark 
        & \nmark & \nmark & \nmark 
        & \nmark & \nmark & \nmark 
        & \nmark & \high & \high \\
    \hline
    Location Reference 
        & \nmark & \vhigh & \nmark 
        & \nmark & \med & \nmark 
        & \nmark & \nmark & \nmark 
        & \nmark & \high & \high \\
    \hline
    Odometry 
        & \nmark & \vhigh & \nmark 
        & \nmark & \high & \nmark 
        & \nmark & \nmark & \nmark 
        & \nmark & \high & \high \\
    \hline
    Time 
        & \nmark & \high & \nmark 
        & \nmark & \med & \nmark 
        & \nmark & \nmark & \nmark 
        & \nmark & \high & \high \\
    \hline
    Emergency Stop 
         & \nmark & \vhigh & \vhigh 
        & \nmark & \high & \high 
        & \nmark & \nmark & \nmark 
        & \nmark & \high & \nmark \\
    \hline
    Acknowledgements 
        & \nmark & \nmark & \nmark 
        & \nmark & \nmark & \nmark 
        & \nmark & \nmark & \nmark 
        & \nmark & \high & \high \\
    \hline
    Train Data 
        & \nmark & \high & \nmark 
        & \nmark & \med & \nmark 
        & \nmark & \nmark & \nmark 
        & \nmark & \high & \high \\
    \hline
    Train Status Reports 
        & \nmark & \nmark & \nmark 
        & \nmark & \nmark & \nmark 
        & \nmark & \nmark & \nmark 
        & \nmark & \high & \high \\
    \hline
    Shunting 
         & \nmark & \nmark & \nmark 
        & \nmark & \nmark & \nmark 
        & \nmark & \nmark & \nmark 
        & \nmark & \med & \low \\
    \hline
    End of Information 
        & \nmark & \nmark & \nmark 
        & \nmark & \nmark & \nmark 
        & \nmark & \nmark & \nmark 
        & \nmark & \nmark & \nmark \\
    \hline
    Error Reporting 
         & \nmark & \med & \nmark 
        & \nmark & \nmark & \nmark 
        & \nmark & \nmark & \nmark 
        & \nmark & \high & \nmark \\
    \hline
    \gls{dmi} Button Press 
        & \nmark & \nmark & \nmark 
        & \nmark & \nmark & \nmark 
        & \nmark & \nmark & \nmark 
        & \nmark & \high & \high \\
    \hline
    \gls{bmm}*
        & \nmark & \nmark & \nmark 
        & \nmark & \nmark & \nmark 
        & \nmark & \nmark & \nmark 
        & \nmark & \high & \high \\
    \hline
    Safe Radio Supervision*
        & \nmark & \med & \med 
        & \nmark & \nmark & \nmark 
        & \nmark & \nmark & \nmark 
        & \nmark & \high & \high \\
    \hline
    \end{tabular}
\end{table}
\section{Controls and Threats}
\label{sec:controls}

This section presents the threats (\Cref{sec:threats-basic,sec:threats-advanced}) and controls (\Cref{sec:controls-basic,sec:controls-advanced}) identified through our methodology, progressing from basic to advanced.
Threats are organized by the underlying technology or communication interface identified in the system model (\Cref{sec:model}). For each technology of a connection and component, \gls{stride} categories are applied (\Cref{sec:methodology}, step 4), informed by known vulnerabilities from the literature. Social engineering is a cross-cutting vector that affects \gls{oob} key management procedures.
\glsxtrshort{rap} values follow \gls{cem}~\cite{cem-2022}; unless stated otherwise, repeated attacks on cryptographic primitives reduce the \gls{rap} to \textit{Basic}, since broken keys can be reused.

\subsection{Basic Threats}
\label{sec:threats-basic}

\textbf{Magnetic Induction (EuroBalise and EuroLoop):} The easiest technology to threaten in \gls{ertms} is the magnetic induction used by EuroBalise and EuroLoop telegrams. 
Disruption of this communication requires only makeshift equipment \cite{railway-cybersecurity-survey-2023}, and long stretches of the track lack physical protection, resulting in a \textit{Basic} \gls{rap}.
To spoof telegrams, custom equipment on the tracks can imitate a balise, though an expert level of knowledge and bespoke equipment may be necessary. 
Furthermore, the attacker may need to spend extended time around the railroad, which could raise suspicion. 
The \gls{rap} rating of this threat is \textit{Moderate}.
An alternative spoofing approach involves replay attacks as suggested by Lim et al. \cite{related-ertms-2019-lim}. 
We consider this to be as difficult as the fake balise threat, yielding a \textit{Moderate} \gls{rap}.
Another concern raised by Lim et al. \cite{related-ertms-2019-lim} is firmware modification.
Although the specification states that such ``test interfaces'' are ``not required to be integrated in the operational equipment'' \cite{subset036}, we still consider firmware reflashing in our analysis. 
Both physical debug and remote interfaces can be used, creating two potential threats. 
In either case, bespoke tools and expert knowledge may be required due to vendor-specific equipment. Both threats receive a \textit{High} \gls{rap}.
Information disclosure is not analyzed here, since telegram data are not secret and their effects can be observed externally.

\noindent\textbf{Ethernet (On-Board Communication):} Breaking the availability of an Ethernet conduit is trivial; severing the cable prevents data flow. This is done in negligible time, though physical presence is required, resulting in a \textit{Basic} \gls{rap}.
Sniffing can be executed with similar difficulty. 
However, specialized equipment may be necessary to reconnect the cable while allowing data interception. The \gls{rap} remains \textit{Basic}.
Lastly, we consider data integrity. Spoofing on an Ethernet network has the same prerequisites as sniffing, since a physical device must be patched into the network. We raise the expertise level to proficient, since knowledge of layer-2 spoofing is required.
However, the \gls{rap} does not increase beyond \textit{Basic}. We do not separately evaluate tampering, since combining the availability and spoofing threats achieves the same goal.

\noindent\textbf{\texorpdfstring{\glsfirst{ip}}{IP}}: 
\gls{ip} spoofing is trivial and ranks lowest in each category, earning a \textit{Basic} \gls{rap} rating.
Again, we do not deem it necessary to analyze tampering attacks separately.
Compromising the confidentiality of \gls{ip} traffic via sniffing requires an on-path position. 
For our purposes, we consider a malicious network infrastructure operator. 
We assume it takes about a week for an attacker with expert knowledge to influence routing protocols in the desired way. The knowledge of \gls{toe} should be assumed to be restricted at least, since the \gls{ip} addresses and ports used are not publicly documented. The necessary equipment should be at least specialized, if not bespoke. The total \gls{rap} equates to \textit{Moderate}.
Disrupting IP traffic availability can be achieved either by an on-path attacker dropping packets (analogous to the sniffing threat) or via large-scale \gls{ddos} attacks. The emergence of ``\gls{ddos} as a Service'' providers enables automated botnet-based flooding for monetary compensation \cite{botnet-ddos-2015}, requiring only restricted TOE knowledge. 
The resulting \gls{rap} is \textit{Enhanced-Basic}, potentially lower if cost estimates of \$10/h are accurate \cite{black-market-report-2018}.

\noindent\textbf{\texorpdfstring{\glsfirst{gsmr}}{GSM-R}:} Due to the prevalence of \gls{gsm} in mobile communication, equipment for sniffing, jamming, and spoofing is readily available. The knowledge of the \gls{toe} ranges from public to restricted, depending on the details, such as the phone numbers used. Each of these three threats can be executed ad hoc if the attacker is adjacent to the track. The resulting \gls{rap} is \textit{Basic} irrespective of the expertise level.

\noindent\textbf{\texorpdfstring{\glsfirst{gnss}}{GNSS}:} Morong et al. \cite{gnss-jamming-2019} demonstrated the ease of \gls{gnss} jamming, requiring only a few watts of transmission power, and calculated the jam range to be tens of kilometers, requiring no specialized equipment, no sensitive knowledge, and no time investment. The resulting \gls{rap} is \textit{Basic}. 
Attacking the integrity of \gls{gnss} is more involved, as correctly spoofing the transmission requires expert understanding of the involved mathematics and signal theory \cite{gnss-spoofing-2016}. 
Furthermore, given more advanced receivers capable of discerning the direction of the incoming signal, the positioning of the antennas must be more deliberate, and the equipment must be at least specialized. 
Still, the instantaneous long-range nature of the threat means the \gls{rap} does not exceed \textit{Enhanced-Basic}. 
Tampering with the \gls{gnss} signal or the satellites themselves is infeasible or overly complex compared to the jam-and-spoof approach. 
Information disclosure is not considered, since \gls{gnss} data is public.

\noindent\textbf{Social Engineering:} 
In general, the adversary might need to be hired by the respective company to gain physical access to the train (estimated time: 1 month). 
Since details of the \gls{km} procedures at the firm are needed, the required \gls{toe} knowledge is restricted. Similarly, the window of opportunity is ranked as difficult. The total \gls{rap} is \textit{High}. For repeated attacks, the time required is reduced to under a day, since the person on the inside only needs to get through the hiring process once, reducing the \gls{rap} to \textit{Moderate}.

\subsection{Basic Controls}
\label{sec:controls-basic}

\textbf{Controls Inherent to the Model:} To simplify the model (\Cref{sec:model}), some of the system properties are expressed as controls. 
Firstly, the lack of \gls{ato} can be viewed as a control, since functions associated with automatic driving cannot be exploited when they are not implemented. This control raises the \gls{rap} of threats related to \gls{ato} to \textit{Beyond High}. Similarly, using \gls{etcs} exclusively in level 2 removes the EuroLoop and the \gls{riu} from the model and limits data transmitted by the EuroBalise to location references \cite{subset026}.
Lastly, we model \gls{frmcs} as a control on \gls{gsmr} technology, since it fully replaces the legacy standard. The control protects against threats to confidentiality and integrity by significantly increasing the time, expertise, and resources required to break state-of-the-art cryptographic primitives, resulting in a \gls{rap} of \textit{Beyond High} \cite{related-ertms-2025-benjumea}.\\
\noindent\textbf{Mandatory Controls:} The \gls{ertms} specifications contain some mandatory protective measures. At the \gls{ob} Ethernet level, the PROFINET standard is used. 
However, it does not introduce any cybersecurity measures and instead relies on ``defense in depth''. 
For our use case, we consider a properly separated network, as recommended by PROFINET and \gls{ertms} \cite{subset147}. We assume that all safety-relevant Ethernet cables are physically difficult for normal passengers to access, thereby increasing the window of opportunity to difficult \cite{profinet-niemann-2019}.
Nearly all communication over \gls{ip}, like between the train and the \gls{pki}, is TLS-secured. 
We assume this has the same effect on \gls{ip} as \gls{frmcs} does on \gls{gsmr} \cite{subset125}. 
One notable exception is \gls{dns}, which uses unencrypted \gls{udp} datagrams.
Lastly, \gls{gsmr} extends the \gls{osi} network model by a ``safety layer''. 
Despite its wording, this layer is responsible for traffic encryption, and the \gls{3des} primitive used in this layer is considered outdated (see \Cref{sec:background-gsmr}) \cite{entropy-nist-2020,transition-nist-2020,3des-sweet32-2016,related-ertms-2015-lopez}.
For our analysis, we estimate that the safety layer increases the \gls{rap} of threats to \gls{gsmr}'s confidentiality and integrity to \textit{Enhanced-Basic}.\\
\noindent\textbf{Optional Controls:} Two optional safety-related concepts in \gls{ertms} apply to cybersecurity. 
Firstly, balise linking, described in \Cref{sec:background-etcs}, can be used as a countermeasure against threats to telegram availability. While it does not change the \gls{rap}, the control \textit{transforms} safety-related damages into train delays through induced emergency halts \cite{subset026}.
Note that balise linking has no effect on other integrity attacks like modified firmware or spoofing of balises.
Secondly, the \gls{ob}'s ``safe radio supervision'' monitors unannounced interruptions of \gls{ota} communication. 
Upon loss of signal, the train either halts or fails to respond, depending on the configuration \cite{subset026}.
Hence, we consider setting the reaction to a trip (or normal braking) as a control on \gls{gsmr}'s availability with the same transformative property as balise linking. 
Moreover, the \gls{ob} Ethernet network specification mentions that ``security functions are expected on higher layers to ensure end-to-end security'' \cite{subset147}, though it is not clear how these are to be implemented while maintaining interoperability. We model this as a control that raises the \gls{rap} of confidentiality and integrity threats on Ethernet to \textit{Beyond High}.

\subsection{Further Controls}
\label{sec:controls-advanced}

\textbf{Existing Literature:} Plenty of research suggests various forms of jamming detection and the use of an \gls{ids} \cite{related-ertms-2025-benjumea,related-ertms-2022-koop,related-ertms-2019-lim,related-ertms-2015-lopez,related-ertms-2023-soderi}. Regrettably, these recommendations tend to be vague, especially regarding the system's response to an alarm. For this reason, we choose not to model these controls.
The proposal by Lim et al. to introduce \glspl{mac} to the telegram format offers a unique approach to hardening the interface without breaking backward compatibility \cite{related-ertms-2019-lim}. They suggest reusing the 12-bit scrambling mechanism \cite{subset036} (a shift-register seeded by a short random value) using derived keys from a shared secret to protect the confidentiality and integrity of balise telegrams \cite{related-ertms-2019-lim}.
While the proposal is creative, we have doubts about its resilience against cyberattacks. Firstly, the short length of the \gls{mac}s allows for computationally feasible collisions. Forging telegrams can still prove challenging, as the scrambling can be seen as encryption. However, the telegrams are very short, and scrambling was never intended for this use. Additionally, the plaintext could be inferred by observing the rolling stock's reaction. 
Based on these concerns, we calculate the \gls{rap} increase of threats to telegram confidentiality and integrity to \textit{Enhanced-Basic}.\\
\label{sec:controls-advanced-original}
\noindent\textbf{Original Suggestions:} Based on early risk indicators, we have composed a few simple controls against threats seen in \Cref{sec:threats-basic}. 
The relatively low security of \gls{oob} key management, balise debug interfaces, and \gls{gnss} leads us to believe that disabling these mechanisms should be considered. 
Prohibition of ``off-line'' key management \cite{subset038} eliminates all discussed social engineering threats. 
Disabling physical and remote debug interfaces protects balise firmware from modification. 
Accurate \gls{ob} time-keeping without reliance on \gls{gnss} mitigates attacks on this system.
Protected \glspl{lx} should serve as a second layer of defense in case of \gls{etcs} compromise. Transferring the responsibility to stop at \glspl{lx} to road vehicles eliminates the danger of trains not being able to halt there in time.
Since the TLS control does not apply to \gls{dns}, its integrity remains vulnerable, leading to massive delays in the best case and advanced impersonation attacks in the worst case. 
We suggest using \gls{dnssec} to eliminate this threat.
Lastly, we believe that balise linking, in its current state, is easily circumvented by fake balises and replay attacks. We propose extending this measure to support an (authenticated) \gls{doe}. In short, the train should react to unannounced balises in the same way as to missed balises, e.g., by tripping. This strategy protects against threats to the integrity of telegrams (other than firmware modification) by transforming safety-related damage scenarios into delay-related ones.
\subsection{Further Threats} Considering all controls, we identify one novel threat: balise linking can be broken by triggering the \gls{bmm} track condition. This is possible through attacks on the integrity of telegrams or \gls{gsmr}, but one can also provoke its legitimate issuance, e.g., by scattering large metallic masses along the tracks that interfere with the magnetic field without affecting drivability. We model this as spoofing a \gls{bmm} datum originating from the \gls{dmi}. The \gls{rap} is \textit{Moderate}, since the attacker needs extended presence at the track and precise calibration of the amount of metal.
\label{sec:threats-advanced}
Other controls are already affected by threats from \Cref{sec:threats-basic}, which we reflect by assigning the affected controls to the functions' security goals, so that their compromise disables the controls. 
We identified three such relationships. Breaking the confidentiality or integrity of \gls{cmp} breaks the controls TLS, \gls{dnssec}, and \gls{frmcs}. Similarly, breaking the confidentiality or integrity of \gls{km} breaks the \gls{gsmr} safety layer. The control balise linking depends on the availability or integrity of \gls{vbc}.
There are other functions whose compromise can serve as an entry point to more sophisticated attacks, which we leave for future research (see \cref{sec:conclusion}).

\section{Evaluation}
\label{sec:evaluation}
Through our methodology, we identified a total of $191$ risks. Appendix \ref{app:risks} shows the complete list of risks with different control groups applied.
We focus on the most substantial ones.
With the most basic control set, i.e., only \Cref{sec:controls-basic} and no \gls{ato}, $17$ risks have a \textit{Low} risk level, $44$ \textit{Moderate}, $62$ \textit{High}, and $68$ \textit{Very High}. 
Much of the \textit{Very High} category stems from the insecurity of EuroBalises, since various forms of spoofing, tampering, and jamming of this technology may easily lead to fatalities. Threats to \gls{gsmr} and \gls{ob} Ethernet buses pose the same danger, except for the edge case of precisely timed cable cuts, which we consider unrealistic.
\Gls{dos} attacks on the \gls{ip} layer are also ranked as \textit{Very High} due to the potential to temporarily halt the entire rail service, e.g., in a scenario where a \gls{kmc} or the \gls{pki} becomes unavailable. A similar effect can be hypothesized for spoofing and tampering of \gls{gnss}. 
Tampering may also lead to life-threatening injuries, should the \gls{ob} clock drift enough to significantly overrun an \gls{loa}.

When all optional \gls{ertms} controls from \Cref{sec:controls-basic}, including \gls{frmcs}, are employed, the numbers of risks in each category shift to $17$/$82$/$44$/$48$, respectively. However, only risks related to \gls{ob} Ethernet and \gls{gsmr} spoofing are fully mitigated\footnote{Raw output suggests that \gls{gsmr} spoofing is possible if \gls{ip} sniffing is done beforehand, as the former breaks TLS encryption and the latter breaks the \gls{gsmr} safety layer. This is a shortcoming of the tooling, which omits dependencies (here, circular) between attack steps, so we ignore this risk.}, i.e., reduced to \textit{Moderate} risk. While balise linking and safe radio supervision prevent jamming attacks from causing severe injuries, they instead result in service disruptions, since the controls respond by applying brakes (damage transformation from safety to operational damages). The combination of low \gls{rap} and potential line-wide disruptions means the effective risk level remains unchanged. Lastly, tampering of telegrams, attacks on \gls{gnss}, and \gls{ddos} remain entirely unaddressed. The reader is reminded that the decrease in \textit{Very High} risk attacks from $68$ to $48$ does not imply full mitigation of the underlying threats, since multiple attack trees may share the same root threat.

Introducing suggestions from \Cref{sec:controls-advanced} shifts the classification to $17$/$89$/$58$/$27$. 
Only attacks on \gls{gnss} are fully mitigated, which explains the low increase of the \textit{Moderate} category's cardinality. 
The three main threats to telegram integrity (replay attacks, fake balises, and firmware reflashing) remain possible (albeit with an increased \gls{rap}). 
However, their damage is transformed from train collisions into delays and cancellations, much as discussed for jamming attacks. 
Due to the non-trivial \gls{rap}, the relevant risk level is reduced from \textit{Very High} to \textit{High}. 
The jamming and \gls{ddos} attacks remain problematic at the \textit{Very High} level.

\begin{figure}[tb]
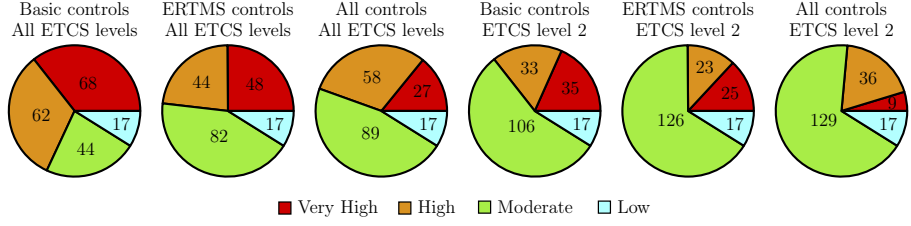

    \centering
    \figurePies
    \caption{Total risk count by risk level per active control group set}
    \label{fig:pies}
\end{figure}

The three control groups are analyzed once more with the inclusion of an ``\gls{etcs} level 2 only'' control, simulating a complete phase-out of levels $0$, \gls{ntc}, and 1. With only mandatory controls, the risk counts are $17$/$106$/$33$/$35$. Including the optional controls in the specifications, we obtain $17$/$126$/$23$/$25$. Finally, enabling all controls yields $17$/$129$/$36$/$9$. In each case, the stark risk reduction is due to the reduced role of EuroBalises. Still, they are used for train positioning, possibly enabling signal and buffer stop overruns that result in severe injuries by gradually shifting the \gls{ob}'s determined location. This is classified as \textit{Very High}. Implementing the linking \gls{doe} measure suggested in \Cref{sec:controls-advanced-original} reduces the risk to \textit{High} by transforming the damages to train delays. The $9$ remaining \textit{Very High} risks refer to previously discussed jamming and \gls{ddos} attacks, which can disrupt the service with relative ease. A visualization of the effects of the various control groups on the risk taxonomy is shown in \Cref{fig:pies}.

Lastly, we consider the activation of \gls{ato}. This is only valid in conjunction with \gls{frmcs}, since \gls{ato} over \gls{gsmr} is not allowed. In each case, we observe only one additional \textit{High} risk level attack, namely cutting the Ethernet cable between the \gls{ato} \gls{obu} and the vehicle interface at just the right moment to inhibit the deactivation of traction. Realistically, this is not concerning, as the \gls{etcs} \gls{obu}'s braking system overrides \gls{ato}, which our model does not capture \cite{subset125}. 
The \gls{ato}'s low effect on the risks stems from the fact that, if an attacker compromises \gls{ob} bus connections or the \gls{frmcs} network, they can cause catastrophic damage by compromising train brakes even without \gls{ato}. 
With the inclusion of \gls{ato}, the attacker could additionally apply full throttle, but that does not change the risk level. 
Furthermore, the attack surface on \gls{ato} is smaller, as all its data is encrypted by TLS.

\section{Discussion}
\label{sec:discussion}
The results show that a minimal \gls{ertms} implementation has several vulnerabilities that can lead to fatalities. 
While additional safety measures defined in the standards, including the upcoming \gls{frmcs}, appear to have a significantly positive effect on the estimated security, many concerns remain regarding \gls{dos} and EuroBalise tampering. 
Additional controls from other researchers and this paper address these deficiencies by strengthening wireless interfaces. 
Nevertheless, the lack of security in telegrams remains a persistent problem. 
Restricting to \gls{etcs} level 2 results in a drastic security improvement by eliminating most EuroBalises' responsibilities, yet attacks on \gls{gnss} and telegram forgery remain unaddressed.
The \gls{ertms} standard appears to be the weakest link in its cybersecurity defenses, which we attribute to the slow pace of \gls{ot} system standardization.
Even with academic countermeasures, the system's reliability remains vulnerable to remote \gls{dos} attacks\textemdash consistent with real-life attacks on railway infrastructure.
Further defenses must be added to ensure uninterrupted operation by avoiding single points of failure using redundant communication, decentralizing control systems, and enabling \gls{v2v} communication.

\section{Conclusion and Future Work}
\label{sec:conclusion}
In this paper, we have performed a risk analysis of the \gls{ertms} suite of standards by systematically translating the specification documents into an abstract system model and subsequently estimating risks using the \gls{mora} framework. By comparing the risk profiles of current and future \gls{ertms} configurations, we identified legacy components, particularly EuroBalises and GSM-R, as the primary source of risk. Fully transitioning to \gls{etcs} level 2 and deploying \gls{frmcs} significantly improves the cybersecurity posture. Jamming and \gls{ddos} remain among the highest risks due to low \gls{rap}s.

\gls{mora} allows flexible handling of \gls{ertms}'s complexity due to its iterative and modular nature and ensures completeness with respect to the model and selected threats, minimizing subjectivity and bias. This work combines function-level security goals for robustness and defensibility of results with technology-based, asset-level threats to achieve complete coverage.

We have identified several limitations in applying \gls{mora} to the railway domain:
\textbf{(1)} Identified risks are classified as having the highest risk level due to the direct effect that train operations have on human safety. \textbf{(2)} Damage transformation (e.g., converting safety hazards into delays) often fails to lower the risk classification when the \gls{rap} of the attacks is relatively low. \textbf{(3)} Some risks are only reduced to the \textit{Moderate} level, even if the corresponding attacks are determined to be no longer possible. \textbf{(4)} Summarizing by count of attack trees may inflate threats with many paths. \textbf{(5)} The lack of temporal dependencies between attack-tree nodes renders circular dependencies unresolvable.

A railway-centric risk matrix addresses the points \textbf{(1)}--\textbf{(3)} as a natural advancement. Further extensions include expanding the model beyond \gls{ertms}, increasing the granularity of specific subsystems (e.g., the attack surface of \gls{frmcs}), and validating identified risks through experiments. Results of the risk assessment can be used by the industry to advance \gls{ertms} cybersecurity.

\begin{credits}
\subsubsection{\ackname} 
Thanks to Patrick Wagner for his advice and review and to Daniel Angermeier for the \gls{mora} tooling (both from Fraunhofer AISEC).

\subsubsection{\discintname}
The authors have no competing interests.

\end{credits}

\bibliographystyle{splncs04}
\bibliography{literature_arXiv}
\newpage

\appendix

\section{Complete List of Risks}
\label{app:risks}
\input{table_without_id}
\section{List of Abbreviations}
\begingroup
\let\clearpage\relax
\let\cleardoublepage\relax
\makeatletter
\printnoidxglossary[type=\acronymtype,title=]
\makeatother
\endgroup

\end{document}

%% file: acronyms.tex
\setabbreviationstyle[acronym]{long-short}
\glssetcategoryattribute{acronym}{nohyperfirst}{true}

\newacronym{ato}{ATO}{Automatic Train Operation}
\newacronym{etcs}{ETCS}{European Train Control System}
\newacronym{ertms}{ERTMS}{European Rail Traffic Management System}
\newacronym{frmcs}{FRMCS}{Future Railway Mobile Communication System}
\newacronym{gsmr}{GSM-R}{\gls{gsm}-Railway}
\newacronym{gsm}{GSM}{Global System for Mobile Communications}
\newacronym{mitm}{MitM}{Man in the Middle}
\newacronym{pki}{PKI}{Public Key Infrastructure}
\newacronym{dns}{DNS}{Domain Name System}
\newacronym{era}{ERA}{European Union Agency for Railways}
\newacronym{ota}{OTA}{Over-the-air}
\newacronym{dos}{DoS}{Denial of Service}
\newacronym{prng}{PRNG}{Pseudo Random Number Generator}
\newacronym{soc}{SOC}{Security Operations Centre}
\newacronym{mdm}{MDM}{Maintenance and Data Management}
\newacronym{cbtc}{CBTC}{Communication Based Train Control}
\newacronym{iv}{IV}{Intra-Vehicular}
\newacronym{ai}{AI}{Artificial Intelligence}
\newacronym{ot}{OT}{Operational Technology}
\newacronym{cc}{CC}{Common Criteria for Information Technology Security Evaluation}
\newacronym{cem}{CEM}{Common Methodology for Information Technology Security Evaluation}
\newacronym{sil}{SIL}{Safety Integrity Levels}
\newacronym{toe}{TOE}{Target of Evaluation}
\newacronym{fwi}{FWI}{Fatalities and Weighted Injuries}
\newacronym{ob}{OB}{On-Board}
\newacronym{obu}{OBU}{On-Board Unit}
\newacronym{ts}{TS}{Track-Side}
\newacronym{ntc}{NTC}{National Train Control}
\newacronym{stm}{STM}{Specific Transmission Module}
\newacronym{rbc}{RBC}{Radio Block Centre}
\newacronym{riu}{RIU}{Radio Infill Unit}
\newacronym{elc}{ELC}{Euroloop Leaky Cable}
\newacronym{ma}{MA}{Movement Authority}
\newacronym{eoa}{EOA}{End of Authority}
\newacronym{loa}{LOA}{Limit of Authority}
\newacronym{dmi}{DMI}{Driver Machine Interface}
\newacronym{gnss}{GNSS}{Global Navigation Satellite System}
\newacronym{tcp}{TCP}{Transmission Control Protocol}
\newacronym{ip}{IP}{Internet Protocol}
\newacronym{cs}{CS}{Circuit Switching}
\newacronym{ps}{PS}{Packet Switching}
\newacronym{3des}{3DES}{Triple Data Encryption Algorithm}
\newacronym{tls}{TLS}{Transport Layer Security}
\newacronym{sp}{SP}{Segment Profile}
\newacronym{jp}{JP}{Journey Profile}
\newacronym{ttsm}{TTSM}{Time Table Speed Management}
\newacronym{ssem}{SSEM}{Supervised Speed Envelope Management}
\newacronym{atsm}{ATSM}{Automa\-tic Train Stopping Management}
\newacronym{km}{KM}{Key Management}
\newacronym{kmc}{KMC}{Key Management Centre}
\newacronym{oob}{OOB}{Out-of-Band}
\newacronym{cbc}{CBC}{Cipher Block Chaining}
\newacronym{mac}{MAC}{Message Authentication Code}
\newacronym{c2m2}{C2M2}{Cybersecurity Capability Maturity Model}
\newacronym{lte}{LTE}{Long-Term Evolution}
\newacronym{dpi}{DPI}{Deep Packet Inspection}
\newacronym{ids}{IDS}{Intrusion Detection System}
\newacronym{cvss}{CVSS}{Common Vulnerability Scoring System}
\newacronym{it}{IT}{Information Technology}
\newacronym{api}{API}{Application Programming Interface}
\newacronym{stride}{STRIDE}{Spoofing, Tampering, Repudiation, Information Disclosure, Denial of Service, Elevation of Privilege}
\newacronym{ord}{ORD}{On-Board Recording Device}
\newacronym{mvb}{MVB}{Multifunction Vehicle Bus}
\newacronym{can}{CAN}{Controller Area Network}
\newacronym{leu}{LEU}{Lineside Electronic Unit}
\newacronym{udp}{UDP}{User Datagram Protocol}
\newacronym{rap}{RAP}{Required Attack Potential}
\newacronym{ddos}{DDoS}{Distributed Denial of Service}
\newacronym{osi}{OSI}{Open Systems Interconnection}
\newacronym{dnssec}{DNSSEC}{Domain Name System Security Extensions}
\newacronym{doe}{DoE}{Denial of Existence}
\newacronym{bmm}{BMM}{Big Metallic Masses}
\newacronym{cmp}{CMP}{Certificate Management Protocol}
\newacronym{vbc}{VBC}{Virtual Balise Cover}
\newacronym{eoi}{EOI}{End of Information}
\newacronym{nist}{NIST}{National Institute of Standards and Technology}
\newacronym{lx}{LX}{level crossing}
\newacronym{v2v}{V2V}{Vehicle-to-Vehicle}
\newacronym{mora}{MoRA}{Modular Risk Assessment for the Development of Secure Automotive Systems}

%% file: figures.tex
\usepackage{pgf-pie}

\newcommand{\pieCol}{
}

\newcommand{\modelCol}{
  \definecolor{sat}{Hsb}{0,1,1}
  \definecolor{obborder}{Hsb}{0,.8,.5}
  \definecolor{gnborder}{Hsb}{40,.8,.5}
  \definecolor{tsborder}{Hsb}{70,.8,.5}
  \definecolor{kmborder}{Hsb}{180,.8,.5}
  \definecolor{inborder}{Hsb}{240,.8,.5}
  \definecolor{obfill}{Hsb}{0,.1,1}
  \definecolor{gnfill}{Hsb}{40,.1,1}
  \definecolor{tsfill}{Hsb}{70,.1,1}
  \definecolor{kmfill}{Hsb}{180,.1,1}
  \definecolor{infill}{Hsb}{240,.1,1}
}

\newcommand{\figurePies}{
    \newsavebox\legendVHi
    \begin{lrbox}{\legendVHi}
      \begin{tikzpicture}
        \large
        \pieCol
        \tikzset{every label/.style={rotate=0}}
        \pie[
          sum=auto,
          color={veryhigh!100},
          text=legend,
          radius=0,
          hide number
        ]{
          9/Very High
        }
      \end{tikzpicture}
    \end{lrbox}
    \newsavebox\legendHi
    \begin{lrbox}{\legendHi}
      \begin{tikzpicture}
        \large
        \pieCol
        \tikzset{every label/.style={rotate=0}}
        \pie[
          sum=auto,
          color={high!100},
          text=legend,
          radius=0,
          hide number
        ]{
          36/High
        }
      \end{tikzpicture}
    \end{lrbox}
    \newsavebox\legendMo
    \begin{lrbox}{\legendMo}
      \begin{tikzpicture}
        \large
        \pieCol
        \tikzset{every label/.style={rotate=0}}
        \pie[
          sum=auto,
          color={low!100},
          text=legend,
          radius=0,
          hide number
        ]{
          129/Moderate
        }
      \end{tikzpicture}
    \end{lrbox}
    \newsavebox\legendLo
    \begin{lrbox}{\legendLo}
      \begin{tikzpicture}
        \large
        \pieCol
        \tikzset{every label/.style={rotate=0}}
        \pie[
          sum=auto,
          color={vlow!100},
          text=legend,
          radius=0,
          hide number
        ]{
          17/Low
        }
      \end{tikzpicture}
    \end{lrbox}
    \newsavebox\legend
    \begin{lrbox}{\legend}
      \begin{tikzpicture}[scale=1,transform shape]
        \node at (1, 2) {\usebox\legendVHi};
        \node at (3.25, 2) {\usebox\legendHi};
        \node at (5.5, 2) {\usebox\legendMo};
        \node at (7.75, 2) {\usebox\legendLo};
      \end{tikzpicture}
    \end{lrbox}
    \begin{tikzpicture}[scale=0.579,transform shape,every text node part/.style={align=center}]
        \large
        \pieCol
        \node at (5, 2) {\usebox\legend};

        \node at (-3.5, 6.35) {Basic controls\\ All \gls{etcs} levels};
        \pie[
          sum=auto,
          color={veryhigh!100, high!100, low!100, vlow!100},
          radius=1.5,
          pos={-3.5,4.25}
        ]{
          68/,
          62/,
          44/,
          17/
        }

        \node at (0, 6.35) {\gls{ertms} controls\\ All \gls{etcs} levels};
        \pie[
          sum=auto,
          color={veryhigh!100, high!100, low!100, vlow!100},
          radius=1.5,
          pos={0,4.25}
        ]{
          48/,
          44/,
          82/,
          17/
        }

        \node at (3.5, 6.35) {All controls\\ All \gls{etcs} levels};
        \pie[
          sum=auto,
          color={veryhigh!100, high!100, low!100, vlow!100},
          radius=1.5,
          pos={3.5,4.25}
        ]{
          27/,
          58/,
          89/,
          17/
        }

        \node at (7, 6.35) {Basic controls\\ \gls{etcs} level $2$};
        \pie[
          sum=auto,
          color={veryhigh!100, high!100, low!100, vlow!100},
          radius=1.5,
          pos={7,4.25}
        ]{
          35/,
          33/,
          106/,
          17/
        }

        \node at (10.5, 6.35) {\gls{ertms} controls\\ \gls{etcs} level $2$};
        \pie[
          sum=auto,
          color={veryhigh!100, high!100, low!100, vlow!100},
          radius=1.5,
          pos={10.5,4.25}
        ]{
          25/,
          23/,
          126/,
          17/
        }

        \node at (14, 6.35) {All controls\\ \gls{etcs} level $2$};
        \pie[
          sum=auto,
          color={veryhigh!100, high!100, low!100, vlow!100},
          radius=1.5,
          pos={14,4.25}
        ]{
          9/,
          36/,
          129/,
          17/
        }
    \end{tikzpicture}
}

\newcommand{\figureModel}{
    \pgfdeclarelayer{bg}
    \pgfsetlayers{bg,main}
    \modelCol
    \begin{tikzpicture}[
      scale=0.73,
      transform shape,
      node distance=1.5cm,
      arr/.style = {semithick,latex-latex},
      ar/.style = {semithick,-latex},
      ra/.style = {semithick,latex-},
      a/.style = {semithick},
    ]
      \tikzset{component/.append style={draw, fill=sat, fill opacity=0.05, text opacity=1, blend group=saturation}}

      \node[component, minimum width=5.5cm, minimum height=3.5cm, name path=etcsborder] (etcs) {\glsxtrshort{etcs} \glsxtrshort{obu}};
      \node[component, minimum width=2cm, minimum height=1cm, left=of etcs, yshift=1.25cm] (stm) {\glsxtrshort{stm}};
      \node[component, minimum width=2cm, minimum height=1cm, below=.25cm of stm] (ord) {\glsxtrshort{ord}};
      \node[component, minimum width=2cm, minimum height=6cm, below=.25cm of ord, name path=vehborder] (veh) {\shortstack{Vehicle\\Interface}};
      \node[component, minimum width=2cm, minimum height=3.5cm, below=of etcs, xshift=1.75cm, name path=radborder] (rad) {EuroRadio};
      \node[component, minimum width=2cm, minimum height=1cm, below=of etcs, xshift=-1.75cm] (dmi) {\glsxtrshort{dmi}};
      \node[component, minimum width=2cm, minimum height=1cm, below=of dmi] (ato) {\glsxtrshort{ato} \glsxtrshort{obu}};

      \node[component, minimum width=2cm, minimum height=1cm, right=5cm of dmi] (rbc) {\glsxtrshort{riu}};
      \node[component, minimum width=2cm, minimum height=1cm, below=.5cm of rbc] (riu) {\glsxtrshort{rbc}};
      \node[component, minimum width=2cm, minimum height=1cm, above=.25cm of rbc] (loop) {EuroLoop};
      \node[component, minimum width=2cm, minimum height=1cm, above=.25cm of loop] (bal) {EuroBalise};
      
      \node[component,minimum width=2cm, minimum height=1cm, below=of rad] (gw) {Gateway};
      \node[component,minimum width=2cm, minimum height=1cm, right=of gw] (pki) {\glsxtrshort{pki}};
      \node[component,minimum width=2cm, minimum height=1cm, below=.5 of pki, xshift=-1.75cm] (atots) {\glsxtrshort{ato} \glsxtrshort{ts}};
      \node[component,minimum width=2cm, minimum height=1cm, left=of gw] (dns) {\glsxtrshort{dns}};

      \node[component, minimum width=2cm, minimum height=2.5cm, right=of riu, yshift=.75cm, name path=kmcborder] (kmc) {\glsxtrshort{kmc}};
      \node[component, minimum width=2cm, minimum height=1cm, right=1.5 of pki] (kmc2) {\shortstack{External\\\glsxtrshort{kmc}}};

      \node[component, minimum width=2cm, minimum height=1cm, right=of etcs, yshift=1.25cm] (gnss) {\glsxtrshort{gnss}};

      \begin{pgfonlayer}{bg}
        \draw[obborder, fill=obfill] (stm.north west)+(-.25cm,+.25cm) rectangle ($(rad.south east)+(.25cm,-.25cm)$);
        \draw[tsborder, fill=tsfill] (bal.north west)+(-.25cm,+.25cm) rectangle ($(riu.south east)+(.25cm,-.25cm)$);
        \draw[kmborder, fill=kmfill] (kmc.north west)+(-.25cm,+.25cm) rectangle ($(kmc2.south east)+(.25cm,-.25cm)$);
        \draw[inborder, fill=infill] (dns.north west)+(-.25cm,+.25cm) rectangle ($(atots.south east)+(1.75cm,0cm)+(.25cm,-.25cm)$);
        \draw[gnborder, fill=gnfill] (gnss.north west)+(-.25cm,+.25cm) rectangle ($(gnss.south east)+(.25cm,-.25cm)$);
      \end{pgfonlayer}{bg}

      \tikzset{every node/.style={sloped, above, midway}}

      \path[name path=proj] (ord.east) -- ($(ord.east)+(2cm,0cm)$);
      \path[ar, name intersections={of=proj and etcsborder}] (intersection-1) edge node {Eth} (ord.east);

      \path[name path=proj] (stm.east) -- ($(stm.east)+(2cm,0cm)$);
      \path[arr, name intersections={of=proj and etcsborder}] (intersection-1) edge node {Eth} (stm.east);

      \path[name path=proj] (dmi.north) -- ($(dmi.north)+(0,2cm)$);
      \path[arr, name intersections={of=proj and etcsborder}] (intersection-1) edge node {Eth} (dmi.north);

      \path[name path=proj] (rad.north) -- ($(rad.north)+(0,2cm)$);
      \path[arr, name intersections={of=proj and etcsborder}] (intersection-1) edge node[rotate=180] {Eth} (rad.north);

      \path[name path=proj] (ato.east) -- ($(ato.east)+(2cm,0cm)$);
      \path[arr, name intersections={of=proj and radborder}] ($(intersection-1)+(0cm,-0.25cm)$) edge node {Eth} ($(ato.east)+(0cm,-0.25cm)$);

      \path[name path=proj] (gnss.west) -- ($(gnss.west)+(-2cm,0cm)$);
      \path[ra, name intersections={of=proj and etcsborder}] (intersection-1) edge node {\glsxtrshort{gnss}} (gnss.west);

      \path[name path=proj] (bal.west) -- ($(bal.west)+(-2cm,0cm)$);
      \path[ra, name intersections={of=proj and etcsborder}] (intersection-1) edge node {Ind} (bal.west);

      \draw[ra] ($(etcs.east)+(0cm,-1.5cm)$) -|- ($(loop.west)+(0,0cm)$) node [midway, above] {Ind};

      \path[name path=proj] (rbc.west) -- ($(rbc.west)+(-2cm,0cm)$);
      \path[ra, name intersections={of=proj and radborder}] (intersection-1) edge node {\glsxtrshort{gsm}} (rbc.west);

      \path[name path=proj] (riu.west) -- ($(riu.west)+(-2cm,0cm)$);
      \path[arr, name intersections={of=proj and radborder}] (intersection-1) edge node {\glsxtrshort{gsm}} (riu.west);

      \path[name path=proj] ($(etcs.south west)+(0cm,.5cm)$) -- ($(etcs.south west)+(-2cm,.5cm)$);
      \path[arr, name intersections={of=proj and vehborder}] (intersection-1) edge node {Eth} ($(etcs.south west)+(0cm,.5cm)$);

      \path[name path=proj] (ato.west) -- ($(ato.west)+(-2cm,0cm)$);
      \path[arr, name intersections={of=proj and vehborder}] (intersection-1) edge node {Eth} (ato.west);

      \path[name path=proj] (rbc.east) -- ($(rbc.east)+(2cm,0cm)$);
      \path[arr, name intersections={of=proj and kmcborder}] ($(intersection-1)+(0,.25cm)$) edge node {\glsxtrshort{tcp}} ($(rbc.east)+(0,.25cm)$);
      \path[arr, name intersections={of=proj and kmcborder}] ($(intersection-1)-(0,.25cm)$) edge node {\glsxtrshort{oob}} ($(rbc.east)-(0,.25cm)$);

      \path[name path=proj] (riu.east) -- ($(riu.east)+(2cm,0cm)$);
      \path[arr, name intersections={of=proj and kmcborder}] ($(intersection-1)+(0,.25cm)$) edge node {\glsxtrshort{tcp}} ($(riu.east)+(0,.25cm)$);
      \path[arr, name intersections={of=proj and kmcborder}] ($(intersection-1)-(0,.25cm)$) edge node {\glsxtrshort{oob}} ($(riu.east)-(0,.25cm)$);

      \draw[arr] ($(ato.east)+(0cm,0.25cm)$) -| (etcs.south) node [near end, above, rotate=180] {Eth};

      \draw[arr] ($(kmc.north)+(-.25cm,0cm)$) |- ($(etcs.east)+(0cm,0cm)+(0cm,-.25cm)$) node [above, sloped=false, pos=0.6] {\glsxtrshort[hyper=false]{oob}};
      \draw[arr] ($(kmc.north)+(.25cm,0cm)$) |- ($(etcs.east)+(0cm,0cm)+(0cm,.25cm)$) node [above, sloped=false, pos=0.6, xshift=-.4cm] {\glsxtrshort[hyper=false]{tcp}};

      \draw[a] ($(pki.north)+(-.5cm,0cm)$) -- ($(riu.south)+(-.5cm,-.5cm)$) node [rotate=180, xshift=-.25cm] {\glsxtrshort[hyper=false]{tcp}} ;
      \draw[ra] ($(pki.north)+(-.5cm,0cm)$) |- ($(riu.south west)+(-.125cm,-.125cm)$) -- ($(riu.south west)+(-.125cm,.25cm)$);
      \draw[dotted] ($(riu.south west)+(-.125cm,.25cm)$) -- ($(riu.north west)+(-.125cm,-.25cm)$);
      \draw[ar] ($(riu.north west)+(-.125cm,-.25cm)$) -- ($(riu.north west)+(-.125cm,.125cm)$) -| (rbc.south);

      \draw[arr] (pki.north) -- (riu.south) node [rotate=180] {\glsxtrshort[hyper=false]{tcp}};

      \draw[arr] (atots.west) -| (gw.south) node [near start, above] {\glsxtrshort{tcp}};

      \draw[arr] (atots.east) -| (pki.south) node [near start, above] {\glsxtrshort{tcp}};

      \draw[arr] ($(pki.north)+(.5cm,0cm)$) |-| ($(kmc.south)+(-.5cm,0cm)$) node [midway, above] {\glsxtrshort{tcp}};

      \draw[arr] ($(kmc2.north)+(0cm,0cm)$) |-| ($(kmc.south)+(0,0cm)$) node [midway, above, rotate=270] {\glsxtrshort[hyper=false]{oob}};
      \draw[arr] ($(kmc2.north)+(.5cm,0cm)$) |-| ($(kmc.south)+(.5,0cm)$) node [midway, above, rotate=270] {\glsxtrshort[hyper=false]{tcp}};

      \path[arr]
        (rad)
          edge node[rotate=180] {\glsxtrshort[hyper=false]{gsm}} (gw)
        
        (dmi)
          edge node[rotate=180] {Eth} (ato)

        (gw)
          edge node {\glsxtrshort{tcp}} (pki)
          edge node {\glsxtrshort{udp}} (dns)
        
        (pki)
          edge node {\glsxtrshort{tcp}} (kmc2)
      ;
    \end{tikzpicture}
}

%% file: table_without_id.tex
\renewcommand{\med}{\cellcolor{low} M}
\renewcommand{\low}{\cellcolor{vlow} L}
{\scriptsize
\begin{longtable}{|p{0.35\textwidth}|p{0.3\textwidth}|p{0.12\textwidth}|p{0.12\textwidth}|p{0.12\textwidth}|}
    \caption{List of all risks (\legendsquare{veryhigh}\textit{\underline{V}ery \underline{H}igh}, \legendsquare{high}\textit{\underline{H}igh}, \legendsquare{low}\textit{\underline{M}oderate}, \legendsquare{vlow}\textit{\underline{L}ow}) with associated threats and different active control groups}
    \label{tab:risks}
    \\
    \hline
    \textbf{Threat} & \textbf{Preperation Threats} & \textbf{Max risk Basic controls} & \textbf{Max risk \gls{ertms} controls} & \textbf{Max risk All controls} \\
    \hline \endhead
Jamming of Balises / EuroLoops &  & \vhigh & \vhigh & \vhigh \\ 
 \hline 
Jamming of Balises / EuroLoops &  Forge a balise & \vhigh & \vhigh & \vhigh \\ 
 \hline 
Jamming of Balises / EuroLoops &  Replay attack on Balises / EuroLoops & \vhigh & \vhigh & \vhigh \\ 
 \hline 
Jamming of Balises / EuroLoops &  Modifying balise firmware remotely & \high & \high & \high \\ 
 \hline 
Jamming of Balises / EuroLoops &  Modifying balise firmware via physical debug connector & \high & \high & \high \\ 
 \hline 
Jamming of Balises / EuroLoops &  Physically force a big metallic mass condition & \vhigh & \med & \med \\ 
 \hline 
Jamming of Balises / EuroLoops &  Spoof GSM-R & \vhigh & \med & \med \\ 
 \hline 
Jamming of Balises / EuroLoops &  Social engineering, Spoof GSM-R & \high & \med & \med \\ 
 \hline 
Jamming of Balises / EuroLoops &  Break physical security on-board a train, Sniffing on an ethernet connection (without AD), Spoof GSM-R & \med & \med & \med \\ 
 \hline 
Jamming of Balises / EuroLoops &  Sniff internet traffic (without AD), Spoof GSM-R & \vhigh & \vhigh & \vhigh \\ 
 \hline 
Jamming of Balises / EuroLoops &  Break physical security on-board a train, Spoof ethernet traffic (without AD) & \med & \med & \med \\ 
 \hline 
Replay attack on Balises / EuroLoops &  & \vhigh & \vhigh & \high \\ 
 \hline 
Replay attack on Balises / EuroLoops &  Forge a balise & \vhigh & \vhigh & \vhigh \\ 
 \hline 
Replay attack on Balises / EuroLoops &  Jamming of Balises / EuroLoops & \vhigh & \vhigh & \high \\ 
 \hline 
Replay attack on Balises / EuroLoops &  Modifying balise firmware remotely & \high & \high & \high \\ 
 \hline 
Replay attack on Balises / EuroLoops &  Modifying balise firmware via physical debug connector & \high & \high & \high \\ 
 \hline 
Replay attack on Balises / EuroLoops &  Physically force a big metallic mass condition & \high & \med & \med \\ 
 \hline 
Replay attack on Balises / EuroLoops &  Spoof GSM-R & \vhigh & \med & \med \\ 
 \hline 
Replay attack on Balises / EuroLoops &  Social engineering, Spoof GSM-R & \med & \med & \med \\ 
 \hline 
Replay attack on Balises / EuroLoops &  Break physical security on-board a train, Sniffing on an ethernet connection (without AD), Spoof GSM-R & \med & \med & \med \\ 
 \hline 
Replay attack on Balises / EuroLoops &  Sniff internet traffic (without AD), Spoof GSM-R & \high & \high & \high \\ 
 \hline 
Replay attack on Balises / EuroLoops &  Break physical security on-board a train, Spoof ethernet traffic (without AD) & \med & \med & \med \\ 
 \hline 
Modifying balise firmware remotely &  Use a debug connection (physical or remote) on a Balise/Euroloop & \high & \high & \med \\ 
 \hline 
Modifying balise firmware via physical debug connector &  Use a debug connection (physical or remote) on a Balise/Euroloop & \high & \high & \med \\ 
 \hline 
Forge a balise &  Pseudo-threat to exclude CTRL\_IND\_MAC from replay attacks on Balises/EuroLoops  & \vhigh & \vhigh & \high \\ 
 \hline 
Forge a balise &  Replay attack on Balises / EuroLoops, Pseudo-threat to exclude CTRL\_IND\_MAC from replay attacks on Balises/EuroLoops  & \vhigh & \vhigh & \high \\ 
 \hline 
Forge a balise &  Jamming of Balises / EuroLoops, Pseudo-threat to exclude CTRL\_IND\_MAC from replay attacks on Balises/EuroLoops  & \vhigh & \vhigh & \high \\ 
 \hline 
Forge a balise &  Modifying balise firmware remotely, Pseudo-threat to exclude CTRL\_IND\_MAC from replay attacks on Balises/EuroLoops  & \high & \high & \med \\ 
 \hline 
Forge a balise &  Modifying balise firmware via physical debug connector, Pseudo-threat to exclude CTRL\_IND\_MAC from replay attacks on Balises/EuroLoops  & \high & \high & \med \\ 
 \hline 
Forge a balise &  Physically force a big metallic mass condition, Pseudo-threat to exclude CTRL\_IND\_MAC from replay attacks on Balises/EuroLoops  & \high & \med & \med \\ 
 \hline 
Forge a balise &  Spoof GSM-R, Pseudo-threat to exclude CTRL\_IND\_MAC from replay attacks on Balises/EuroLoops  & \vhigh & \med & \med \\ 
 \hline 
Forge a balise &  Social engineering, Spoof GSM-R, Pseudo-threat to exclude CTRL\_IND\_MAC from replay attacks on Balises/EuroLoops  & \med & \med & \med \\ 
 \hline 
Forge a balise &  Break physical security on-board a train, Sniffing on an ethernet connection (without AD), Spoof GSM-R, Pseudo-threat to exclude CTRL\_IND\_MAC from replay attacks on Balises/EuroLoops  & \med & \med & \med \\ 
 \hline 
Forge a balise &  Sniff internet traffic (without AD), Spoof GSM-R, Pseudo-threat to exclude CTRL\_IND\_MAC from replay attacks on Balises/EuroLoops  & \high & \high & \high \\ 
 \hline 
Forge a balise &  Break physical security on-board a train, Spoof ethernet traffic (without AD), Pseudo-threat to exclude CTRL\_IND\_MAC from replay attacks on Balises/EuroLoops  & \med & \med & \med \\ 
 \hline 
Physically force a big metallic mass condition &  & \high & \med & \med \\ 
 \hline 
Sniffing on an ethernet connection &  Break physical security on-board a train & \low & \low & \low \\ 
 \hline 
Cut traffic on an ethernet connection &  Break physical security on-board a train & \med & \med & \med \\ 
 \hline 
Spoof ethernet traffic &  Break physical security on-board a train & \med & \med & \med \\ 
 \hline 
Spoof internet traffic &  & \med & \med & \med \\ 
 \hline 
Spoof internet traffic &  Break physical security on-board a train, Sniffing on an ethernet connection & \med & \med & \med \\ 
 \hline 
Sniff internet traffic &  & \low & \low & \low \\ 
 \hline 
Sniff internet traffic &  Break physical security on-board a train, Sniffing on an ethernet connection & \low & \low & \low \\ 
 \hline 
Drop internet traffic &  & \med & \med & \med \\ 
 \hline 
Social engineering &  & \high & \high & \med \\ 
 \hline 
Jam GSM-R &  & \vhigh & \vhigh & \vhigh \\ 
 \hline 
Jam GSM-R &  Jamming of Balises / EuroLoops & \vhigh & \vhigh & \vhigh \\ 
 \hline 
Jam GSM-R &  Forge a balise, Jamming of Balises / EuroLoops & \vhigh & \vhigh & \vhigh \\ 
 \hline 
Jam GSM-R &  Replay attack on Balises / EuroLoops, Jamming of Balises / EuroLoops & \vhigh & \vhigh & \vhigh \\ 
 \hline 
Jam GSM-R &  Modifying balise firmware remotely, Jamming of Balises / EuroLoops & \high & \high & \high \\ 
 \hline 
Jam GSM-R &  Modifying balise firmware via physical debug connector, Jamming of Balises / EuroLoops & \high & \high & \high \\ 
 \hline 
Jam GSM-R &  Physically force a big metallic mass condition, Jamming of Balises / EuroLoops & \vhigh & \med & \med \\ 
 \hline 
Jam GSM-R &  Spoof GSM-R, Jamming of Balises / EuroLoops & \vhigh & \med & \med \\ 
 \hline 
Jam GSM-R &  Social engineering, Spoof GSM-R, Jamming of Balises / EuroLoops & \high & \med & \med \\ 
 \hline 
Jam GSM-R &  Break physical security on-board a train, Sniffing on an ethernet connection (without AD), Spoof GSM-R, Jamming of Balises / EuroLoops & \med & \med & \med \\ 
 \hline 
Jam GSM-R &  Sniff internet traffic (without AD), Spoof GSM-R, Jamming of Balises / EuroLoops & \vhigh & \vhigh & \vhigh \\ 
 \hline 
Jam GSM-R &  Break physical security on-board a train, Spoof ethernet traffic (without AD), Jamming of Balises / EuroLoops & \med & \med & \med \\ 
 \hline 
Jam GSM-R &  Replay attack on Balises / EuroLoops & \vhigh & \vhigh & \vhigh \\ 
 \hline 
Jam GSM-R &  Forge a balise, Replay attack on Balises / EuroLoops & \vhigh & \vhigh & \vhigh \\ 
 \hline 
Jam GSM-R &  Jamming of Balises / EuroLoops, Replay attack on Balises / EuroLoops & \vhigh & \vhigh & \vhigh \\ 
 \hline 
Jam GSM-R &  Modifying balise firmware remotely, Replay attack on Balises / EuroLoops & \high & \high & \high \\ 
 \hline 
Jam GSM-R &  Modifying balise firmware via physical debug connector, Replay attack on Balises / EuroLoops & \high & \high & \high \\ 
 \hline 
Jam GSM-R &  Physically force a big metallic mass condition, Replay attack on Balises / EuroLoops & \high & \med & \med \\ 
 \hline 
Jam GSM-R &  Spoof GSM-R, Replay attack on Balises / EuroLoops & \vhigh & \med & \med \\ 
 \hline 
Jam GSM-R &  Social engineering, Spoof GSM-R, Replay attack on Balises / EuroLoops & \med & \med & \med \\ 
 \hline 
Jam GSM-R &  Break physical security on-board a train, Sniffing on an ethernet connection (without AD), Spoof GSM-R, Replay attack on Balises / EuroLoops & \med & \med & \med \\ 
 \hline 
Jam GSM-R &  Sniff internet traffic (without AD), Spoof GSM-R, Replay attack on Balises / EuroLoops & \high & \high & \high \\ 
 \hline 
Jam GSM-R &  Break physical security on-board a train, Spoof ethernet traffic (without AD), Replay attack on Balises / EuroLoops & \med & \med & \med \\ 
 \hline 
Jam GSM-R &  Forge a balise & \vhigh & \vhigh & \vhigh \\ 
 \hline 
Jam GSM-R &  Replay attack on Balises / EuroLoops, Forge a balise & \vhigh & \vhigh & \vhigh \\ 
 \hline 
Jam GSM-R &  Jamming of Balises / EuroLoops, Forge a balise & \vhigh & \vhigh & \vhigh \\ 
 \hline 
Jam GSM-R &  Modifying balise firmware remotely, Forge a balise & \high & \high & \high \\ 
 \hline 
Jam GSM-R &  Modifying balise firmware via physical debug connector, Forge a balise & \high & \high & \high \\ 
 \hline 
Jam GSM-R &  Physically force a big metallic mass condition, Forge a balise & \high & \med & \med \\ 
 \hline 
Jam GSM-R &  Spoof GSM-R, Forge a balise & \vhigh & \med & \med \\ 
 \hline 
Jam GSM-R &  Social engineering, Spoof GSM-R, Forge a balise & \med & \med & \med \\ 
 \hline 
Jam GSM-R &  Break physical security on-board a train, Sniffing on an ethernet connection (without AD), Spoof GSM-R, Forge a balise & \med & \med & \med \\ 
 \hline 
Jam GSM-R &  Sniff internet traffic (without AD), Spoof GSM-R, Forge a balise & \high & \high & \high \\ 
 \hline 
Jam GSM-R &  Break physical security on-board a train, Spoof ethernet traffic (without AD), Forge a balise & \med & \med & \med \\ 
 \hline 
Jam GSM-R &  Modifying balise firmware remotely & \high & \high & \high \\ 
 \hline 
Jam GSM-R &  Modifying balise firmware via physical debug connector & \high & \high & \high \\ 
 \hline 
Jam GSM-R &  Spoof GSM-R & \vhigh & \med & \med \\ 
 \hline 
Jam GSM-R &  Social engineering, Spoof GSM-R & \high & \med & \med \\ 
 \hline 
Jam GSM-R &  Break physical security on-board a train, Sniffing on an ethernet connection (without AD), Spoof GSM-R & \med & \med & \med \\ 
 \hline 
Jam GSM-R &  Sniff internet traffic (without AD), Spoof GSM-R & \vhigh & \vhigh & \vhigh \\ 
 \hline 
Sniff GSM-R &  & \low & \low & \low \\ 
 \hline 
Sniff GSM-R &  Social engineering & \low & \low & \low \\ 
 \hline 
Sniff GSM-R &  Break physical security on-board a train, Sniffing on an ethernet connection (without AD) & \low & \low & \low \\ 
 \hline 
Sniff GSM-R &  Sniff internet traffic (without AD) & \low & \low & \low \\ 
 \hline 
Spoof GSM-R &  & \vhigh & \med & \med \\ 
 \hline 
Spoof GSM-R &  Social engineering & \high & \med & \med \\ 
 \hline 
Spoof GSM-R &  Break physical security on-board a train, Sniffing on an ethernet connection (without AD) & \med & \med & \med \\ 
 \hline 
Spoof GSM-R &  Sniff internet traffic (without AD) & \vhigh & \vhigh & \vhigh \\ 
 \hline 
Jam GNSS &  & \vhigh & \vhigh & \med \\ 
 \hline 
Spoof GNSS &  & \vhigh & \vhigh & \med \\ 
 \hline 
Break physical security on-board a train &  & \low & \low & \low \\ 
 \hline 
(Distributed) Denial of Service &  & \med & \med & \med \\ 
 \hline 
Pseudo-threat to exclude CTRL\_IND\_MAC from replay attacks on Balises/EuroLoops  &  & \low & \low & \low \\ 
 \hline 
Use a debug connection (physical or remote) on a Balise/Euroloop &  & \low & \low & \low \\ 
 \hline 
Jamming of Balises / EuroLoops (only ETCS2) &  & \vhigh & \vhigh & \vhigh \\ 
 \hline 
Jamming of Balises / EuroLoops (only ETCS2) &  Forge a balise (only ETCS2) & \high & \high & \high \\ 
 \hline 
Jamming of Balises / EuroLoops (only ETCS2) &  Replay attack on Balises / EuroLoops (only ETCS2) & \high & \high & \high \\ 
 \hline 
Jamming of Balises / EuroLoops (only ETCS2) &  Modifying balise firmware remotely (only ETCS2) & \high & \high & \high \\ 
 \hline 
Jamming of Balises / EuroLoops (only ETCS2) &  Modifying balise firmware via physical debug connector (only ETCS2) & \high & \high & \high \\ 
 \hline 
Jamming of Balises / EuroLoops (only ETCS2) &  Physically force a big metallic mass condition & \high & \med & \med \\ 
 \hline 
Jamming of Balises / EuroLoops (only ETCS2) &  Spoof GSM-R (only ETCS2) & \vhigh & \med & \med \\ 
 \hline 
Jamming of Balises / EuroLoops (only ETCS2) &  Social engineering, Spoof GSM-R (only ETCS2) & \high & \med & \med \\ 
 \hline 
Jamming of Balises / EuroLoops (only ETCS2) &  Break physical security on-board a train, Sniffing on an ethernet connection (without AD), Spoof GSM-R (only ETCS2) & \med & \med & \med \\ 
 \hline 
Jamming of Balises / EuroLoops (only ETCS2) &  Sniff internet traffic (without AD), Spoof GSM-R (only ETCS2) & \high & \high & \high \\ 
 \hline 
Jamming of Balises / EuroLoops (only ETCS2) &  Break physical security on-board a train, Spoof ethernet traffic (without AD) & \med & \med & \med \\ 
 \hline 
Replay attack on Balises / EuroLoops (only ETCS2) &  & \vhigh & \vhigh & \high \\ 
 \hline 
Replay attack on Balises / EuroLoops (only ETCS2) &  Forge a balise (only ETCS2) & \vhigh & \vhigh & \high \\ 
 \hline 
Replay attack on Balises / EuroLoops (only ETCS2) &  Jamming of Balises / EuroLoops (only ETCS2) & \vhigh & \vhigh & \high \\ 
 \hline 
Replay attack on Balises / EuroLoops (only ETCS2) &  Modifying balise firmware remotely (only ETCS2) & \high & \high & \high \\ 
 \hline 
Replay attack on Balises / EuroLoops (only ETCS2) &  Modifying balise firmware via physical debug connector (only ETCS2) & \high & \high & \high \\ 
 \hline 
Replay attack on Balises / EuroLoops (only ETCS2) &  Physically force a big metallic mass condition & \high & \med & \med \\ 
 \hline 
Replay attack on Balises / EuroLoops (only ETCS2) &  Spoof GSM-R (only ETCS2) & \vhigh & \med & \med \\ 
 \hline 
Replay attack on Balises / EuroLoops (only ETCS2) &  Social engineering, Spoof GSM-R (only ETCS2) & \med & \med & \med \\ 
 \hline 
Replay attack on Balises / EuroLoops (only ETCS2) &  Break physical security on-board a train, Sniffing on an ethernet connection (without AD), Spoof GSM-R (only ETCS2) & \med & \med & \med \\ 
 \hline 
Replay attack on Balises / EuroLoops (only ETCS2) &  Sniff internet traffic (without AD), Spoof GSM-R (only ETCS2) & \high & \high & \high \\ 
 \hline 
Replay attack on Balises / EuroLoops (only ETCS2) &  Break physical security on-board a train, Spoof ethernet traffic (without AD) & \med & \med & \med \\ 
 \hline 
Modifying balise firmware remotely (only ETCS2) &  & \high & \high & \high \\ 
 \hline 
Modifying balise firmware via physical debug connector (only ETCS2) &  & \high & \high & \high \\ 
 \hline 
Forge a balise (only ETCS2) &  & \vhigh & \vhigh & \high \\ 
 \hline 
Forge a balise (only ETCS2) &  Replay attack on Balises / EuroLoops (only ETCS2) & \vhigh & \vhigh & \high \\ 
 \hline 
Forge a balise (only ETCS2) &  Jamming of Balises / EuroLoops (only ETCS2) & \vhigh & \vhigh & \high \\ 
 \hline 
Forge a balise (only ETCS2) &  Modifying balise firmware remotely (only ETCS2) & \high & \high & \high \\ 
 \hline 
Forge a balise (only ETCS2) &  Modifying balise firmware via physical debug connector (only ETCS2) & \high & \high & \high \\ 
 \hline 
Forge a balise (only ETCS2) &  Physically force a big metallic mass condition & \high & \med & \med \\ 
 \hline 
Forge a balise (only ETCS2) &  Spoof GSM-R (only ETCS2) & \vhigh & \med & \med \\ 
 \hline 
Forge a balise (only ETCS2) &  Social engineering, Spoof GSM-R (only ETCS2) & \med & \med & \med \\ 
 \hline 
Forge a balise (only ETCS2) &  Break physical security on-board a train, Sniffing on an ethernet connection (without AD), Spoof GSM-R (only ETCS2) & \med & \med & \med \\ 
 \hline 
Forge a balise (only ETCS2) &  Sniff internet traffic (without AD), Spoof GSM-R (only ETCS2) & \high & \high & \high \\ 
 \hline 
Forge a balise (only ETCS2) &  Break physical security on-board a train, Spoof ethernet traffic (without AD) & \med & \med & \med \\ 
 \hline 
Jam GSM-R (only ETCS2) &  & \vhigh & \vhigh & \vhigh \\ 
 \hline 
Jam GSM-R (only ETCS2) &  Jamming of Balises / EuroLoops (only ETCS2) & \vhigh & \vhigh & \vhigh \\ 
 \hline 
Jam GSM-R (only ETCS2) &  Forge a balise (only ETCS2), Jamming of Balises / EuroLoops (only ETCS2) & \vhigh & \vhigh & \high \\ 
 \hline 
Jam GSM-R (only ETCS2) &  Replay attack on Balises / EuroLoops (only ETCS2), Jamming of Balises / EuroLoops (only ETCS2) & \vhigh & \vhigh & \high \\ 
 \hline 
Jam GSM-R (only ETCS2) &  Modifying balise firmware remotely (only ETCS2), Jamming of Balises / EuroLoops (only ETCS2) & \high & \high & \high \\ 
 \hline 
Jam GSM-R (only ETCS2) &  Modifying balise firmware via physical debug connector (only ETCS2), Jamming of Balises / EuroLoops (only ETCS2) & \high & \high & \high \\ 
 \hline 
Jam GSM-R (only ETCS2) &  Physically force a big metallic mass condition, Jamming of Balises / EuroLoops (only ETCS2) & \vhigh & \med & \med \\ 
 \hline 
Jam GSM-R (only ETCS2) &  Spoof GSM-R (only ETCS2), Jamming of Balises / EuroLoops (only ETCS2) & \vhigh & \med & \med \\ 
 \hline 
Jam GSM-R (only ETCS2) &  Social engineering, Spoof GSM-R (only ETCS2), Jamming of Balises / EuroLoops (only ETCS2) & \high & \med & \med \\ 
 \hline 
Jam GSM-R (only ETCS2) &  Break physical security on-board a train, Sniffing on an ethernet connection (without AD), Spoof GSM-R (only ETCS2), Jamming of Balises / EuroLoops (only ETCS2) & \med & \med & \med \\ 
 \hline 
Jam GSM-R (only ETCS2) &  Sniff internet traffic (without AD), Spoof GSM-R (only ETCS2), Jamming of Balises / EuroLoops (only ETCS2) & \vhigh & \vhigh & \vhigh \\ 
 \hline 
Jam GSM-R (only ETCS2) &  Break physical security on-board a train, Spoof ethernet traffic (without AD), Jamming of Balises / EuroLoops (only ETCS2) & \med & \med & \med \\ 
 \hline 
Jam GSM-R (only ETCS2) &  Replay attack on Balises / EuroLoops (only ETCS2) & \vhigh & \vhigh & \high \\ 
 \hline 
Jam GSM-R (only ETCS2) &  Forge a balise (only ETCS2), Replay attack on Balises / EuroLoops (only ETCS2) & \vhigh & \vhigh & \high \\ 
 \hline 
Jam GSM-R (only ETCS2) &  Jamming of Balises / EuroLoops (only ETCS2), Replay attack on Balises / EuroLoops (only ETCS2) & \vhigh & \vhigh & \high \\ 
 \hline 
Jam GSM-R (only ETCS2) &  Modifying balise firmware remotely (only ETCS2), Replay attack on Balises / EuroLoops (only ETCS2) & \high & \high & \high \\ 
 \hline 
Jam GSM-R (only ETCS2) &  Modifying balise firmware via physical debug connector (only ETCS2), Replay attack on Balises / EuroLoops (only ETCS2) & \high & \high & \high \\ 
 \hline 
Jam GSM-R (only ETCS2) &  Physically force a big metallic mass condition, Replay attack on Balises / EuroLoops (only ETCS2) & \high & \med & \med \\ 
 \hline 
Jam GSM-R (only ETCS2) &  Spoof GSM-R (only ETCS2), Replay attack on Balises / EuroLoops (only ETCS2) & \vhigh & \med & \med \\ 
 \hline 
Jam GSM-R (only ETCS2) &  Social engineering, Spoof GSM-R (only ETCS2), Replay attack on Balises / EuroLoops (only ETCS2) & \med & \med & \med \\ 
 \hline 
Jam GSM-R (only ETCS2) &  Break physical security on-board a train, Sniffing on an ethernet connection (without AD), Spoof GSM-R (only ETCS2), Replay attack on Balises / EuroLoops (only ETCS2) & \med & \med & \med \\ 
 \hline 
Jam GSM-R (only ETCS2) &  Sniff internet traffic (without AD), Spoof GSM-R, Replay attack on Balises / EuroLoops (only ETCS2) & \high & \high & \high \\ 
 \hline 
Jam GSM-R (only ETCS2) &  Break physical security on-board a train, Spoof ethernet traffic (without AD), Replay attack on Balises / EuroLoops (only ETCS2) & \med & \med & \med \\ 
 \hline 
Jam GSM-R (only ETCS2) &  Forge a balise (only ETCS2) & \vhigh & \vhigh & \high \\ 
 \hline 
Jam GSM-R (only ETCS2) &  Replay attack on Balises / EuroLoops (only ETCS2), Forge a balise (only ETCS2) & \vhigh & \vhigh & \high \\ 
 \hline 
Jam GSM-R (only ETCS2) &  Jamming of Balises / EuroLoops (only ETCS2), Forge a balise (only ETCS2) & \vhigh & \vhigh & \high \\ 
 \hline 
Jam GSM-R (only ETCS2) &  Modifying balise firmware remotely (only ETCS2), Forge a balise (only ETCS2) & \high & \high & \high \\ 
 \hline 
Jam GSM-R (only ETCS2) &  Modifying balise firmware via physical debug connector (only ETCS2), Forge a balise (only ETCS2) & \high & \high & \high \\ 
 \hline 
Jam GSM-R (only ETCS2) &  Physically force a big metallic mass condition, Forge a balise (only ETCS2) & \high & \med & \med \\ 
 \hline 
Jam GSM-R (only ETCS2) &  Spoof GSM-R (only ETCS2), Forge a balise (only ETCS2) & \vhigh & \med & \med \\ 
 \hline 
Jam GSM-R (only ETCS2) &  Social engineering, Spoof GSM-R (only ETCS2), Forge a balise (only ETCS2) & \med & \med & \med \\ 
 \hline 
Jam GSM-R (only ETCS2) &  Break physical security on-board a train, Sniffing on an ethernet connection (without AD), Spoof GSM-R (only ETCS2), Forge a balise (only ETCS2) & \med & \med & \med \\ 
 \hline 
Jam GSM-R (only ETCS2) &  Sniff internet traffic (without AD), Spoof GSM-R (only ETCS2), Forge a balise (only ETCS2) & \high & \high & \high \\ 
 \hline 
Jam GSM-R (only ETCS2) &  Break physical security on-board a train, Spoof ethernet traffic (without AD), Forge a balise (only ETCS2) & \med & \med & \med \\ 
 \hline 
Jam GSM-R (only ETCS2) &  Modifying balise firmware remotely (only ETCS2) & \high & \high & \high \\ 
 \hline 
Jam GSM-R (only ETCS2) &  Modifying balise firmware via physical debug connector (only ETCS2) & \high & \high & \high \\ 
 \hline 
Jam GSM-R (only ETCS2) &  Spoof GSM-R (only ETCS2) & \vhigh & \med & \med \\ 
 \hline 
Jam GSM-R (only ETCS2) &  Social engineering, Spoof GSM-R (only ETCS2) & \high & \med & \med \\ 
 \hline 
Jam GSM-R (only ETCS2) &  Break physical security on-board a train, Sniffing on an ethernet connection (without AD), Spoof GSM-R (only ETCS2) & \med & \med & \med \\ 
 \hline 
Jam GSM-R (only ETCS2) &  Sniff internet traffic (without AD), Spoof GSM-R (only ETCS2) & \vhigh & \vhigh & \vhigh \\ 
 \hline 
Sniff GSM-R (only ETCS2) &  & \low & \low & \low \\ 
 \hline 
Sniff GSM-R (only ETCS2) &  Social engineering & \low & \low & \low \\ 
 \hline 
Sniff GSM-R (only ETCS2) &  Break physical security on-board a train, Sniffing on an ethernet connection (without AD) & \low & \low & \low \\ 
 \hline 
Sniff GSM-R (only ETCS2) &  Sniff internet traffic (without AD) & \low & \low & \low \\ 
 \hline 
Spoof GSM-R (only ETCS2) &  & \vhigh & \med & \med \\ 
 \hline 
Spoof GSM-R (only ETCS2) &  Social engineering & \high & \med & \med \\ 
 \hline 
Spoof GSM-R (only ETCS2) &  Break physical security on-board a train, Sniffing on an ethernet connection (without AD) & \med & \med & \med \\ 
 \hline 
Spoof GSM-R (only ETCS2) &  Sniff internet traffic (without AD) & \vhigh & \vhigh & \vhigh \\ 
 \hline 
Sniffing on an ethernet connection (without AD) &  & \low & \low & \low \\ 
 \hline 
Cut traffic on an ethernet connection (without AD) &  & \vhigh & \vhigh & \vhigh \\ 
 \hline 
Spoof ethernet traffic (without AD) &  & \vhigh & \med & \med \\ 
 \hline 
Spoof internet traffic (without AD) &  & \med & \med & \med \\ 
 \hline 
Spoof internet traffic (without AD) &  Break physical security on-board a train, Sniffing on an ethernet connection (without AD) & \med & \med & \med \\ 
 \hline 
Sniff internet traffic (without AD) &  & \low & \low & \low \\ 
 \hline 
Sniff internet traffic (without AD) &  Break physical security on-board a train, Sniffing on an ethernet connection (without AD) & \low & \low & \low \\ 
 \hline 
Drop internet traffic (without AD) &  & \vhigh & \vhigh & \vhigh \\ 
 \hline 
(Distributed) Denial of Service (without AD) &  & \vhigh & \vhigh & \vhigh \\ 
 \hline 
\end{longtable}
}